\documentclass{article}

\usepackage[text={173 mm,25cm},centering,letterpaper]{geometry}
\usepackage{amsmath, amsthm, amssymb}
\usepackage{graphicx}
\usepackage[margin = 10pt, font=small, labelfont=bf, labelsep = endash]{caption}
\usepackage[pdfpagemode=UseOutlines, hidelinks]{hyperref}
\usepackage{subcaption}
\usepackage[utf8]{inputenc}

\newcommand{\COO}{CO$_2$}

\newcommand{\cO}{\mathcal{O}}
\newcommand{\cR}{\mathcal{R}}
\DeclareMathOperator{\sech}{sech}

\begin{document}

\title{Climate Response and Sensitivity: Timescales and Late Tipping Points}

\author{Robbin Bastiaansen\thanks{Institute for Marine and Atmospheric research Utrecht, Department of Physics, Utrecht University, The Netherlands (r.bastiaansen@uu.nl, a.s.vonderheydt@uu.nl)} , Peter Ashwin\thanks{Department of Mathematics, University of Exeter, Exeter EX4 4QF, UK (p.ashwin@exeter.ac.uk)} , Anna S. von der Heydt\footnotemark[1]}

\maketitle

\begin{abstract}
Climate response metrics are used to quantify the Earth's climate response to anthropogenic changes of atmospheric \COO. Equilibrium Climate Sensitivity (ECS) is one such metric that measures the equilibrium response to \COO\ doubling. However, both in their estimation and their usage, such metrics make assumptions on the linearity of climate response, although it is known that, especially for larger forcing levels, response can be nonlinear. Such nonlinear responses may become visible immediately in response to a larger perturbation, or may only become apparent after a long transient. In this paper, we illustrate some potential problems and caveats when estimating ECS from transient simulations. We highlight ways that very slow timescales may lead to poor estimation of ECS even if there is seemingly good fit to linear response over moderate timescales. Moreover, such slow timescale might lead to late abrupt responses ("late tipping points") associated with a system's nonlinearities. We illustrate these ideas using simulations on a global energy balance model with dynamic albedo. We also discuss the implications for estimating ECS for global climate models, highlighting that it is likely to remain difficult to make definitive statements about the simulation times needed to reach an equilibrium.
\end{abstract}

\section{Introduction}

The central question as to how the climate is likely to change as a function of anthropogenic \COO\ emissions can be posed as `How does an observation of the climate system respond to changes in its radiative forcing induced by changes in atmospheric \COO?'. This question has been studied in various ways for at least over a century~\cite{Arrhenius1896, aboutArrhenius}, although efforts to answer it became more intense and in-depth over the last decades. Amongst early efforts was the pioneering work by Charney {\em et al} in 1979, who made the first estimates of expected equilibrium warming after doubling of atmospheric \COO\ (while keeping vegetation and land ice fixed at present-day values) using a numerical Global Climate Model (GCM) \cite{Charney.1979}. This metric has later been named the Equilibrium Climate Sensitivity (ECS) and is still widely used. Since then, researchers have developed a number of different metrics that measure climate response to different scenarios of anthropogenic change in \COO\, and have incorporated information from other sources besides computer models, including historical observations and data from palaeoclimate records. Recently, these efforts were summarised in an assessment of the World Climate Research Programme \cite{Sherwood.2020} that synthesised different quantifications of climate response using these different lines lines of evidence and led to the headline that the Earth's ECS is likely between $2.6K$ and $3.9K$.

One of the hurdles for this assessment was the variety of definitions of (the quantification of) climate sensitivity -- and ECS especially -- in the literature. The root of this problem can be attributed to the lack of data on equilibrium climate states or detailed long-term transient data. This can be due to low time resolutions in proxy data, lack of observational data or insufficient computing power to equilibrate modern GCMs. Consequently, equilibrium properties need to be estimated from incomplete data sets, leading to many slightly different ways to quantify climate sensitivity. Common to them all, however, is the need to extrapolate long-term dynamics from data on shorter time scales. In this paper, we describe and discuss this extrapolation process in detail, hereby focusing on estimates of ECS using (idealised) experiments in climate models for the sake of mathematical simplicity. Of particular interest here is the exploration of linear, and non-linear, dynamics that can emerge in multiscale dynamical systems that can cause problems with extrapolation.

The common way to obtain estimates of ECS in climate models involves the use of extrapolation and regression methods on non-equilibrated transient simulations -- typically of 150 year long runs. Values for ECS obtained in this way are now often referred to as the {\it effective climate sensitivity} \cite{senior2000time} signalling that it might not encompass all long-term climate change. Although there are many different ways to perform such extrapolation, common is that it is usually based on linear concepts and frameworks. A recent review \cite{knutti2015feedbacks} of climate sensitivity highlighted that it is a key challenge to study the limits of such linear frameworks. Here, we will investigate these limits and in the process highlight the trade-offs that need to be made when designing experiments to quantify ECS: in order to measure a clear signal of warming in relation to the noise of natural variations, large perturbations are desirable but precisely in the case of larger perturbations the nonlinear behaviour becomes important and linear frameworks break down.

One of the most important tools to study past and future climate change are the GCMs as used in Coupled Model Intercomparison Projects (CMIP, e.g. \cite{eyring2016overview}), because they provide a globally complete and detailed representation of the climate state while (approximately) satisfying the physical laws. However, specifically for these large models there is no way to determine whether a model really has arrived in the linear regime near an equilibrium, or even if such an equilibrium exists. In this paper we explore some simple conceptual examples of the potential nonlinear dynamics of the climate. We also make a number of observations that we hope illuminate some of the limitations of linear frameworks. (i) We highlight cases where there may be strong dependence on the climate background state and the forcing levels. (ii) We highlight examples where there may be a {\em good fit} to transient data but {\em poor extrapolation} preventing an accurate estimation of the ECS. (iii) We show that nonlinear systems can have \emph{slow tipping points}. When these are crossed the tipping dynamics play out on slow time scales, and it can take arbitrarily long times before nonlinear and/or asymptotic behaviour is observed. (iv) We demonstrate how in the presence of multiple-timescales with nonlinear feedbacks a {\em late tipping} can occur in which fast processes suddenly dominate after arbitrarily long slow transient behaviour. This highlights the potential for slow and/or late tipping points to be particular obstructions to estimating ECS.

The rest of this paper is organised as follows: in the remainder of this section we discuss in general the response of a nonlinear system to forcing. In section~\ref{sec:EqResponse}, we consider the equilibrium response and equilibrium climate sensitivity of the climate system in terms of limiting behaviour. Moreover, we point out the challenges that arise when estimating those from short time series, highlighting the trade-offs that emerge in terms of perturbation size and required simulation time. In section~\ref{sec:GEBM}, we examine the nonlinear effects that may appear as a result of climate dynamics on multiple timescales, including {\em slow} tipping which may in turn lead to {\em late but rapid} tipping. We illustrate these effects using  multi-scale global energy balance models with dynamic albedo and/or chaotic variability and an example from a LongRunMIP abrupt8xCO2 run \cite{rugenstein2019longrunmip}. Finally, we briefly discuss these results, and the influence of time-varying forcing on estimation of climate response and sensitivity in Section~\ref{sec:Discuss}.

\subsection{Response of nonlinear models to forcing}

Consider a notional state of the climate system $y(t)$ for $t > t_0$ that evolves in response to various (unknown) forcings, with a partially known initial state $y_0$ and an input of atmospheric \COO\ generating a radiative forcing $\Delta F(t)$ that is specified for $t>t_0$. We write this climate state at time $t$ as 
\begin{equation}
y(t)=Y_t(y_0,t_0,\Delta F),
\end{equation}
where $Y_t$ is an evolution operator that evolves forward the initial state $y_0$ (at time $t_0$) up to time $t$ according to a climate model $Y$ with (possibly time-dependent) radiative forcing $\Delta F$.

Given a scalar observable $\cO$ that maps the full climate state $y(t)$, the response clearly depends on the choice of observable $\cO$, the choice of model $Y$, the forcing $\Delta F$ experienced by the system, the initial climate state $y_0$ at time $t_0$ and the time moment $t>t_0$ of interest.

At the level of a single initial state $y_0$ starting at $t_0$ of which we have perfect knowledge and subject to deterministic forcing $\Delta F$, the response in the observable $\cO$ at time $t>t_0$ is the difference in the observable's value at times $t_0$ and $t$, i.e.
\begin{equation}
\cR_{\cO,Y}\left(t; t_0, y_0; \Delta F\right)= \cO(Y_t(y_0,t_0,\Delta F))-\cO(Y_{t_0}(y_0,t_0,0)).
\label{eq:generic_response}
\end{equation}
This corresponds to a {\em two-point} response in the terminology of \cite{ashwin2020extreme}. As this is often the easiest response type to think about mathematically (and extensions to other types are possible albeit more technical), it is this response type we will be referring to throughout this paper. However, often we are interested not in specific trajectories but rather in the distribution of possible responses for a probability distribution $\mu_0$ of initial states and forcing $\Delta F$. In this case we write the response as
\begin{equation}
\cR_{\cO,Y}\left(t; t_0, \mu_0; \Delta F\right)= \cO(Y_t(\mu_0,t_0,\Delta F))-\cO(Y_{t_0}(\mu_0,t_0,0)).
\label{eq:generic_measure_response}
\end{equation}
This corresponds to a {\em distributional} response, namely it is a random variable with some distribution determined by the ``pushforward'' of the initial probability distribution $\mu_0$ by the dynamics.  
Furthermore, there are different interpretations of (\ref{eq:generic_measure_response}), depending on the choice of probability function. These include:
\begin{itemize}
    \item A {\em physical measure} on a climate attractor \cite{von2016state,ashwin2020extreme}. This can be an observable measured in a long palaeoclimate time series, or an observable in a model, where the attractor is (partly) known from the underlying model equations. 
    \item An {\em ensemble} of initial conditions that are thought to sample subgrid processes in a model (or observational data).
    \item An {\em empirical measure} for a finite segment of trajectory, i.e. a choice of states on $ \{ Y_t(y_0,t_0,0) ~:~t\in[t_0,t_1] \} $  over some finite interval with $t_0<t_1$, with equal weight to any given time instant. Such a measure can be approximated from a finite length time series of a palaeoclimate record.
\end{itemize}
Note that $\mu_0$ is a physical measure means that for typical initial conditions the empirical measures converge to one and the same distribution: for a more precise definition of a physical measure, see for example \cite{eckmann1985ergodic,young2017generalizations}. If there are multiple attractors then there can be several physical measures, and typical initial conditions converge to one of these depending which basin of attraction they are in.

\section{Equilibrium Response and ECS as limiting behaviour}
\label{sec:EqResponse}

While the response on any time scale can be relevant, often the asymptotic, or equilibrium, response as $t \rightarrow \infty$ is considered first. This response is typically easy to analyse and understand in simple models. Taking the limit $t \rightarrow \infty$ of \eqref{eq:generic_response}, the equilibrium response is:
\begin{equation}
    \lim_{t \rightarrow \infty} \mathcal{R}_{O,Y}(t; t_0, y_0; \Delta F).
\end{equation}
Of course, this begs the question of whether the limit exists. In particular, one cannot expect such limit to hold for any forcing $\Delta F$. For instance, if the forcing specifies uninhibited and constant emission of greenhouse gases, the climate system will not evolve to any equilibrium. Hence it makes sense to limit ourselves to forcing scenarios that have constant forcing levels as $t \rightarrow \infty$ (i.e. $\Delta F(t) \rightarrow \Delta F_*$ as $t \rightarrow \infty$). In practical model studies of equilibrium climate sensitivity, often the forcing is just taken as a constant throughout the whole simulation.

Of particular interest is the equilibrium response to an instantaneous and abrupt doubling of atmospheric CO$_2$, which we indicate by the forcing $\Delta F_\mathrm{abrupt2xCO2}$. Then, the equilibrium climate sensitivity (ECS) is defined as the response of global mean surface temperature (GMST) to such forcing, i.e.
\begin{equation}
    \mathrm{ECS}(y_0) := \lim_{t \rightarrow \infty} \mathcal{R}_{\mathrm{GMST},Y}(t; t_0, y_0; \Delta F_\mathrm{abrupt2xCO2}).
    \label{eq:ECS_definition}
\end{equation}

Even for such idealised forcing, such a limit may not be well-defined. In any but the simplest models, the asymptotic climate state will have stationary internal variability, for which the limit of the two-point response is not well-defined without first averaging for long enough that any internal variability is averaged out. In such cases, a distributional response may have a well-defined limit, although it can happen that even these do not converge in cases where there is non-ergodic behaviour \cite{young2002srb}. 

It is difficult to say anything definitive about the convergence of climate response in state-of-the-art GCMs. These models are numerical representations of the underlying physical equations, which have been developed to include many physical processes and ever-improving parametrizations of sub-grid scale processes; they are very high-dimensional and complex. We do not have access to the attractors of these models and so cannot exclude the possibility of poor or no convergence. These models are roughly calibrated only by assessing how well they can reproduce the {\em present day} climate, including the historical period.  However, in practice, reaching the true equilibrium may also be less relevant with such a model; the physical state of the climate systems after a few centuries or even millennia could be difficult to predict anyway because of incomplete knowledge of the initial state $y_0$, model details and forcing. For these reasons, a pragmatic {\em Effective Climate Sensitivity} \cite{senior2000time,Rugenstein2021ThreeFlavors} is often taken, in which response over a few centuries or millennia is taken, ignoring dynamics on longer time scales. However, we focus here on cases where the limit in \eqref{eq:ECS_definition} is well-defined.

\subsection{Background State, Forcing Scenario and ECS}\label{sec:background_state}

In \eqref{eq:ECS_definition}, it is clear that the equilibrium climate sensitivity depends on the initial condition $y_0$ or {\em background state} where the latter refers to the initial climate attractor. However, often ECS is given without explicitly stating initial conditions. This can lead to ambiguity about what is meant by ECS when comparing simulations of current and palaeoclimates. Because of the possibility of multistability of the climate system, even for the same \COO-level, may support multiple climate states. In physical terms, the dependence on the background state originates from feedback processes that changes as the forcing is applied \cite{von2016state}, necessitating a proper communication of the background state considered when computing the ECS of that background state.

Further, in the definition of ECS~\eqref{eq:ECS_definition} a doubling of atmospheric \COO\ is given as forcing scenario. However, in practice, ECS is often used as a measure of temperature increase \emph{per} \COO\ doubling. So by assuming linearity of the climate response to forcing levels, ECS is employed to estimate warming for other \COO\ forcing levels. Specifically, for an abrupt $2^\gamma$x\COO\ forcing, an assumption of linear response would mean that warming of $\gamma$ times the ECS is expected:
\begin{equation}
    \lim_{t \rightarrow \infty} \mathcal{R}_{\mathrm{GMST},Y}(t;t_0,y_0;\Delta F_{\mathrm{abrupt}2^\gamma\mathrm{xCO2}}) = \gamma\ \mathrm{ECS}(y_0).
    \label{eq:ECS_linearity_assumption}
\end{equation}
Certainly, this assumption will fail when $\gamma$ is large enough that a tipping point is crossed, but even when that does not happen such linear assumption only holds when the forcing is small enough that nonlinear terms can be ignored.

It has been shown that this linearity assumption in fact does break down in GCMs. For instance, palaeoclimate simulations with a wide range of \COO-concentrations suggest such linearity can be broken \cite{Caballero.2013} and multi-millennial experiments in the model intercomparison project LongRunMIP~\cite{rugenstein2019longrunmip} also show  deviations from linearity; it was found that abrupt4xCO2 experiments lead to more than twice the warming of an abrupt2xCO2 experiment in the same GCM. Further, abrupt8xCO2 experiments led to less than twice the warming of an abrupt4xCO2 experiment. Hence, the usage of ECS as a linear predictor for warming based on \COO\ levels can easily lead to over- or underestimations of warming.

\subsection{Challenges to estimating ECS from timeseries}
\label{sec:Estimating}

It is computationally expensive to run state-of-the-art GCMs and in principle, millennial length simulations may be needed to get close to equilibrium (see e.g. the LongRunMIP \cite{rugenstein2020equilibrium}). Because there exists variability on many time scales and spatial feedback patterns in these models, there is no {\em a priori} method to determine when or indeed whether a nonlinear model has reached equilibrium. This means that the equilibrium response of a climate model cannot be directly found from time evolution of the model; instead, one needs to derive and extrapolate the equilibrium properties of the model from possibly relatively short transient data.

In general, estimation of ECS for a model (such as a GCM) involves four steps:
\begin{enumerate}
    \item {\em Design} of an experimental protocol (initial conditions, forcing levels, simulation time, ensemble of runs of the GCM);
    \item {\em Selection} of a time period for fitting;
    \item {\em Fitting} of transient observable data to a less complex model;
    \item {\em Extrapolation} to derive equilibrium properties from the fitted model.
\end{enumerate}

Many different protocols have been used -- see e.g. \cite[Table 2]{Rugenstein2021ThreeFlavors} that lists 11 different methodologies. However, the most common standard for estimating ECS uses a technique by Gregory et al \cite{gregory2004new}. Typically, a single abrupt \COO-forcing experiment is run (starting from pre-industrial forcing levels, standard is to use an abrupt 4xCO2 forcing) for some years (150 years is the benchmark for CMIP6 models). The transient data on change in the yearly and globally averaged observables near-surface-temperature $\Delta T$ and top-of-atmosphere radiative imbalance $\Delta N$ is fitted to the linear model $\Delta N = \lambda \Delta T + f$. Then, equilibrium warming $\Delta T^*$ is estimated setting $\Delta N = 0$ in this linear model (since, in equilibrium, there should be radiative balance), yielding $\Delta T^*_\mathrm{est} = - \lambda^{-1} f$. 
Albeit its predominant use in climate sensitivity analyses in GCMs, it is clear that GCMs are not well-approximated by this simple linear model over all time scales; because climate feedback processes operate at quite different timescales, $\Delta N$ and $\Delta T$ will have a non-linear relationship that has non-zero curvature over the course of a long simulation, and the linear relationship only holds approximately for certain time intervals \cite{Bastiaansen:2021fm,rugenstein2019longrunmip, andrews2015dependence, knutti2017beyond}. Better fits to the response over all the time scales can be found by considering a combination of several linearly decaying modes, i.e. by viewing the climate system as a combination of linear processes with quite different time scales \cite{Cummins.2020,Bastiaansen:2021fm, dai2020improved, geoffroy2013transient}. 

Other protocols use results from the literature of linear response theory directly~\cite{Lucarini.2018, ruelle2009review, Ragone.2016, lucarini2011statistical, proistosescu2017slow, hasselmann1993cold, lembo2020beyond, aengenheyster2018point, bastiaansen2021projections}. That is, in relative generality, the response (of an observable $O$) in the linear regime of a (non-linear) system to a forcing can be characterised via a (causal linear observational) Green's function $G^{[O]}(t)$. Specifically, the yearly and globally (and ensemble) average near-surface-temperature increase $\Delta T$ at time $t$ under a certain forcing scenario $\Delta F$ is given by the relation
$$
\Delta T(t) = \left( G^{[T]} \ast \Delta F \right)(t) := \int_0^t G^{[T]}(s) \Delta F(t-s)\ ds.
$$
Using this relationship, transient data can be used to estimate the Green's function $G^{[T]}$ from which the equilibrium response can be extrapolated -- which can be done through fitting to some prescribed function (typically a sum of decaying exponential functions) or through a discrete Fourier transform algorithm.

\begin{figure}
    \centering
    \includegraphics[width=\textwidth]{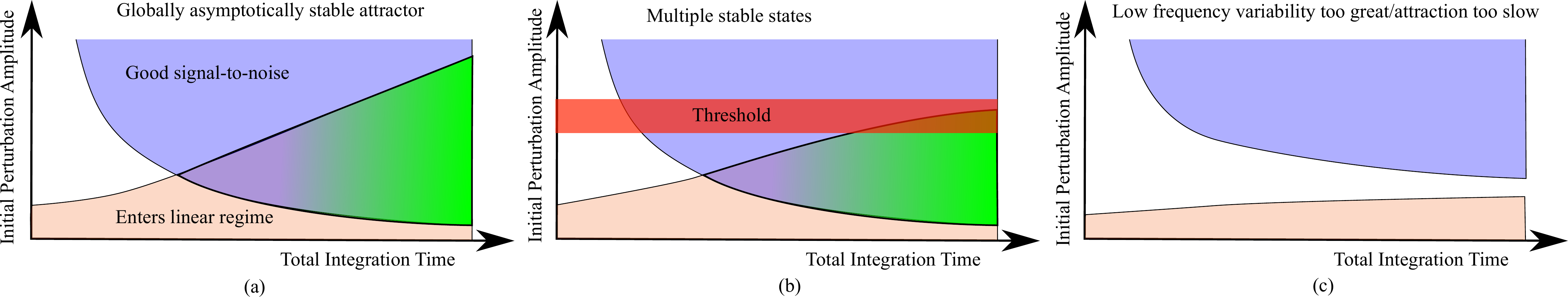}
    \caption{Schematic diagrams illustrating trade-offs between perturbation amplitude and integration time when computing ECS on perturbing a linearly stable state of a nonlinear climate model. The light blue regions illustrate the trade-off needed to give a good signal-to-noise ratio of the estimate of ECS. The pink region illustrates the trade-off needed to ensure the system has entered the linear regime. The green ``Goldilocks zone'' shows points where accurate prediction of ECS is possible. (a) illustrates a case where the state is globally stable while (b) shows a case where a large enough perturbation (above the red bar) pushes the system out of the linear regime - perturbations above this may in principle give super-long transients and/or convergence to another stable state. Finally, (c) shows a case where accurate estimation of ECS is not possible.}
    \label{fig:tradeoff5}
\end{figure}

For all the fitting and extrapolation protocols, the optimal choices in the protocol are not always obvious as certain trade-offs need to be made:
\begin{enumerate}
    \item The simulation time needs to be as long as possible to ensure (a) we are in the linear response of the final equilibrium state and (b) fluctuations caused by natural variability can be averaged out. {\em However,} long simulations for GCMs are computationally expensive and even these will not be able to detect slow timescales beyond the length of simulation time.
    \item A large ensemble and/or a long time period for fitting needs to be chosen to reduce noise caused by internal variability. {\em However,} each additional ensemble member increases the simulation effort and the time period for fitting needs to start as late as possible to maximise the chance of being in a linear regime.
    \item The perturbation needs to be as large as possible to maximise the signal-to-noise ratio for the fitting procedure. {\em However,} large perturbations may result in nonlinear effects, including tipping into different climate states.
\end{enumerate}
Figure~\ref{fig:tradeoff5} illustrates two important trade-offs between perturbation size and integration time. In particular, the figure highlights the need to find a ``Goldilocks Zone'' where the perturbation is neither too small nor too big. Examples of these trade-offs in a nonlinear setting using an conceptual energy balance model are discussed within Section~\ref{sec:GEBM}.

\subsection{Slow linear responses and ECS}

We start by illustrating some challenges that already arise in the linear response regime of a model. In such setting, extrapolation can be difficult if the time scale of the slowest response exceeds the length of timeseries available.  To illustrate this, we now consider the evolution of a linear observable $O$ of a finite $M$-dimensional linear system. In the absence of repeated eigenvalues, the Green's function will be a sum of exponential functions (with exponents being the eigenvalues) with the following functional form:
\begin{equation}
    G^{[O]}(t) = 
    \left\{\begin{array}{cl}
    \sum_{j=1}^M \beta_j^{[O]} e^{\lambda_j t}&\mbox{ if }t\geq 0\\
    0 & \mbox{ if }t<0
    \end{array}\right.
    \label{eq:GreenFunction_assumption}
\end{equation}
where $\lambda_j \in \mathbb{C}$ represent eigenvalues of the linear system and $\beta_j^{[O]} \in \mathbb{R}$ depends on corresponding eigenvector and observable; often $\lambda_j$ is restricted to the negative reals but more generally they may be complex with oscillatory decay (see e.g. \cite{torres2021identification}).

Estimating the Green's function for high (or infinite) dimensional systems can be extremely challenging -- not least because linear operators in infinite dimensions may have a continuous (operator) spectrum. Nonetheless, one can assume a functional form for $G^{[O]}(t)$, and fit parameters from transient data. This approach has been applied successfully to many response problems in the climate system, see e.g.~\cite{bastiaansen2021projections,  torres2021identification, hasselmann1993cold, maier1987transport}. 

\begin{figure}
    \centering
    \includegraphics[width=10cm]{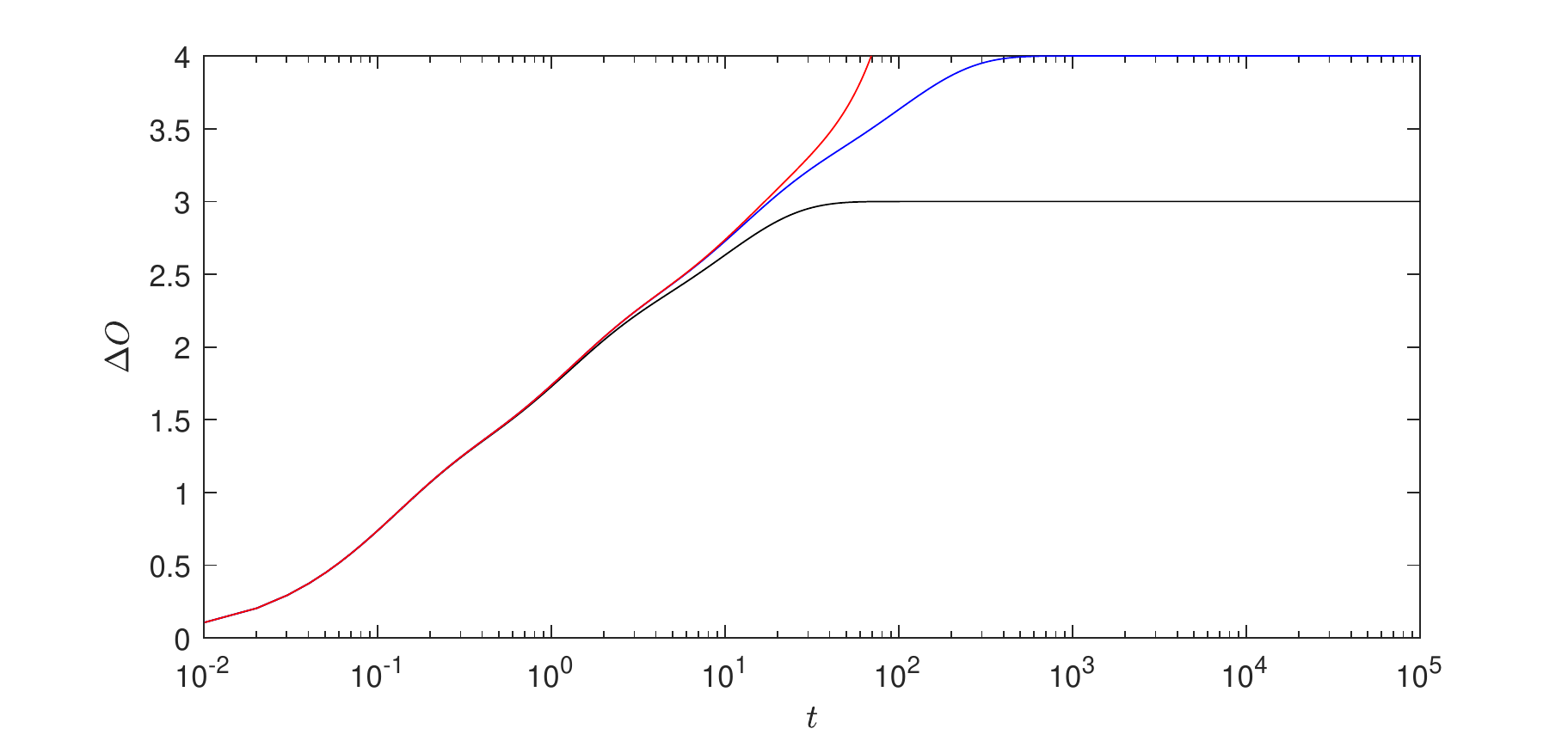}
    \caption{Examples of the response of observables $\Delta O_1$ (black), $\Delta O_2$ (blue) and $\Delta O_3$ (red), sums of exponential functions as defined in \eqref{eq:GFexample}. Note that the $t$-axis is given in log-scale to highlight how long these different equations stay almost indistinguishable: the red case corresponds to a linearly unstable setting, i.e. a `run-away' scenario..}
    \label{fig:GF_exp_example}
\end{figure}

Let us now assume that \eqref{eq:GreenFunction_assumption} holds for the Green's function, and restrict to $\lambda \in \mathbb{R}$. Even then, the number of modes $M$ needs to be determined, and that comes with its own problems as shown in Figure~\ref{fig:GF_exp_example}. This figures compares responses of an observable given by one of the following:
\begin{equation}
\begin{aligned}
\Delta O_1(t) &= 3 - e^{-10t} - e^{-t} - e^{-0.1 t}
\\
\Delta O_2(t) &= 4 - e^{-10t} - e^{-t} - e^{-0.1 t} - e^{-0.01 t}
\\
\Delta O_3(t) &= 2 - e^{-10t} - e^{-t} - e^{-0.1 t} + e^{+0.01 t}
\end{aligned}
\label{eq:GFexample}
\end{equation}
These three examples differ only by the absence/presence of an eigenvalue $\lambda$ with $|\lambda|$ small: only the first two are bounded and these have different asymptotic values; the third describes a `run-away' response. Nonetheless, Figure~\ref{fig:GF_exp_example} shows that all three observables are indistinguishable at first; only over longer time scales does the effect of the small eigenvalue become apparent. It is practically impossible to determine which of these functional forms is correct from short-time transient data only. 

In the climate system, the dynamics play out over many different time scales \cite{anna2020quantification, mitchell1976overview}. Hence, this should play an important role in understanding GCM experiments. In particular, it is important to try to determine time scales on which the constructed estimations and extrapolations can be trusted, as there seems to be no way to completely rule out slow warming, or even slow tipping, on all slow time scales. GCMs very often do not include the very slow climate components such as land ice sheets dynamically, but still need very long spin-up times and almost never are integrated to full equilibrium. For example palaeoclimate experiments with GCMs typically show considerable drifts in the globally averaged ocean temperature after several millennia of simulation, while already in good radiative balance (e.g. \cite{Baatsen:2020vz}).

\section{Nonlinear response and ECS for climate models}
\label{sec:GEBM}

In the previous section, we discussed potential problems associated with timescales that can affect estimation of ECS  even for linear systems. In this section, we turn our attention to issues related to non-linear response. Here, a particular challenge are tipping points where fast dynamics can suddenly take over even after long, very slowly evolving transient periods; this is impossible in a purely linear system.

To make our considerations in this section more explicit, we consider a global energy balance model (GEBM) that has dynamics on two timescales and the possibility of tipping phenomena on a slow or a fast timescale. We introduce the model in subsection (a). Then, we consider tipping-related effects in this model due to time-scale separation of physical processes in subsection (b) or due to internal variability in subsection (c).

\subsection{A fast-slow energy balance model}

We consider a GEBM of Budyko-Sellers-Ghill type~\cite{budyko,sellers,ghil1976climate}, which describes the evolution of GMST $T$ according to the model
\begin{equation}
    C\frac{dT}{dt} = Q_0 (1 - \alpha) - \varepsilon \sigma T^4 + \mu+\mu_{NV}(t),
    \label{eq:0DGEBM}
\end{equation}
where $C$ is the specific heat capacity, $Q_0$ is the incoming (predominantly short wave) solar radiation, $\alpha$ is the planetary albedo (so that $Q_0\alpha$ is the reflected solar radiation) and $\varepsilon\sigma T^4$ is the outgoing (predominantly long-wave) Planck radiation (with planetary emissivity $\varepsilon$ and Boltzmann constant $\sigma$). Further, $\mu$ represents the mean radiative forcing due to increases in \COO\ and $\mu_{NV}(t)$ models variability in radiative forcing, assumed to have zero mean. Following \cite{Myhre2013}, we assume
\begin{equation}
\mu = \mu_0 + A_0 \log\left[\frac{\rho(t)}{\rho(0)}\right]
\label{eq:mu}
\end{equation}
with $A_0 = 5.35 Wm^{-2}$, where $\rho(t)$ is the concentration of atmospheric \COO\ at time $t$, and $\mu_0$ is a reference radiative forcing level for a \COO\ concentration of $\rho(0)$. 

When albedo and/or emissivity are taken to be temperature-dependent, i.e. $\alpha=\alpha(T)$ and/or $\varepsilon = \varepsilon(T)$, the model can have multiple stable climate states each with different climate sensitivity. In this paper we assume there is relaxation towards an equilibrium albedo $\alpha_0(T)$ at a rate $\tau_{\alpha}\geq 0$
\begin{equation}
\tau_{\alpha} \frac{d\alpha}{dt} = \left[ \alpha_0(T) - \alpha \right].
    \label{eq:fsalpha}
\end{equation}
We assume a temperature-dependent equilibrium albedo $\alpha_0(T)$ given by
\begin{equation}
    \alpha_0(T) = \alpha_1 + (\alpha_2 - \alpha_1) \frac{1 + \tanh(K_\alpha [T - T_{\alpha}])}{2}
    \label{eq:alpha0}
\end{equation}
and an instantaneously settling emissivity $\varepsilon(T)$ given by
\begin{equation}
    \varepsilon(T) = \varepsilon_1 + (\varepsilon_2-\varepsilon_1) \frac{1 + \tanh(K_\varepsilon [T - T_\varepsilon])}{2}.
    \label{eq:varepsilon}
\end{equation}
Both of these functional forms are of sigmoid-type, and change from one constant to another as $T$ moves through a range of temperatures near $T_{\alpha,\varepsilon}$ \cite{ashwin2020extreme}; $\alpha_0(T)$ models the (relatively slow) lowering of albedo in the presence of land ice sheets, while $\varepsilon_0(T)$ models a (relatively fast) transition from a clear to a cloudy planet with large quantities of low cloud. Each of them on their own can lead to a bistability between a colder and a warmer climate state but we include both to allow the possibility of independent slow and fast tipping points. In fact, we believe that both the (slowly settling) temperature-dependent albedo and emissivity are required to have some of the later illustrated phenomena -- late tipping in particular -- that do not present themselves in models with constant albedo or emissivity.

We include natural variability of the energy input at the surface represented by chaotic forcing through a Lorenz-63 model, i.e. natural variability $\mu_{NV}$ is given by
\begin{equation}
    \mu_\mathrm{NV} = \nu_\mathrm{NV} \sin(\pi x(t) / 20).
    \label{eq:NV}
\end{equation}
where $x$ adheres to the Lorenz-63 model, which conceptually represents the chaotic dynamics of weather processes \cite{lorenz1963deterministic}
\begin{equation}
	\left\{\begin{aligned}
		\tau_{NV} \frac{dx}{dt}  &=  \sigma (y-x)  \\
		\tau_{NV} \frac{dy}{dt}  &=   x ( \rho - z ) - y  \\
		\tau_{NV} \frac{dz}{dt}  &=   x y - \beta z 
	\end{aligned}\right.
	\label{eq:chaos}
\end{equation}
so that $\nu_\mathrm{NV}$ is a measure for the strength of the variability and $\tau_{NV}$ is the characteristic timescale of chaotic variability. Parameter values used in the simulations in this paper are given in Table~\ref{tab:fsGEBM}, except where stated otherwise.

There are two special parameter settings that we distinguish. We say there is {\em dynamic albedo} if $\tau_{\alpha}>0$; in the case $\tau_{\alpha}=0$, albedo settles instantaneously so that we can eliminate (\ref{eq:fsalpha}) and set $\alpha=\alpha_0(T)$. We say there is {\em chaotic variability} if $\nu_{NV}\neq 0$; in the case $\nu_{NV}=0$, there is no internal variability and we can eliminate the chaotic Lorenz-63 model (\ref{eq:chaos}).

\begin{figure}
    \centering
    \begin{subfigure}[t]{0.35\textwidth}
         \includegraphics[width=\textwidth]{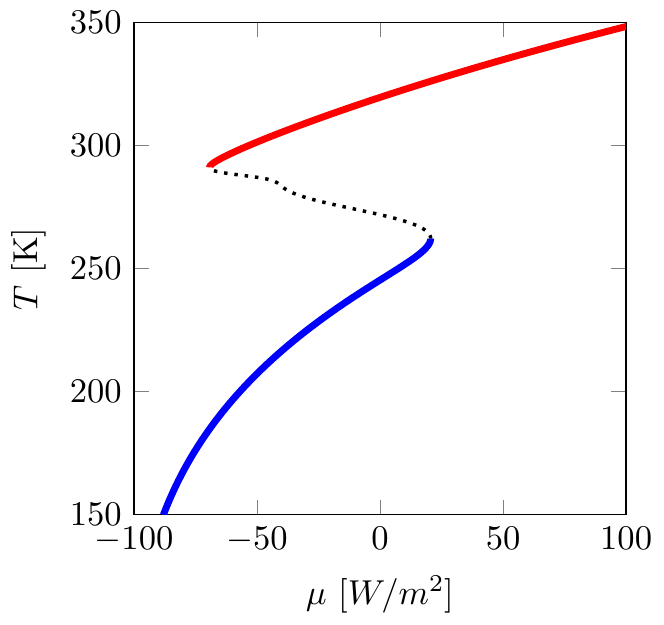}
         \caption{Bifurcation Diagram}
    \end{subfigure}
~
    \begin{subfigure}[t]{0.35\textwidth}
         \includegraphics[width=\textwidth]{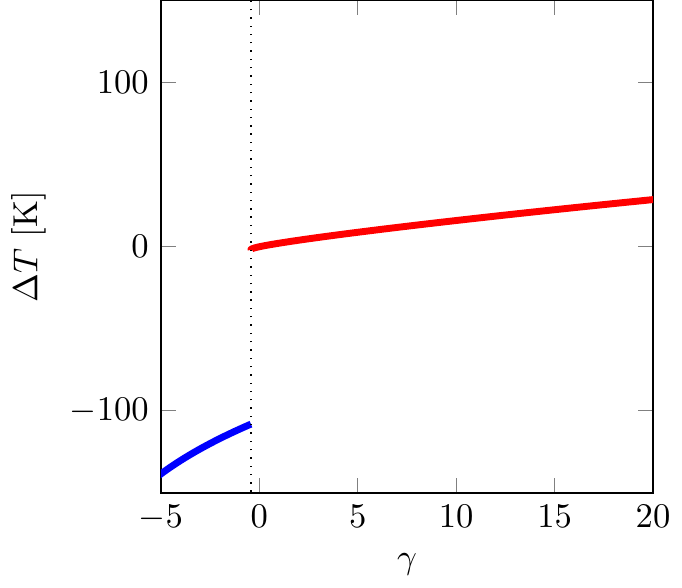}
         \caption{Response for $2^\gamma\times$\COO\ with $T_0 = 293K$.}
    \end{subfigure}
    
~
    \begin{subfigure}[t]{0.7\textwidth}
         \includegraphics[width=\textwidth]{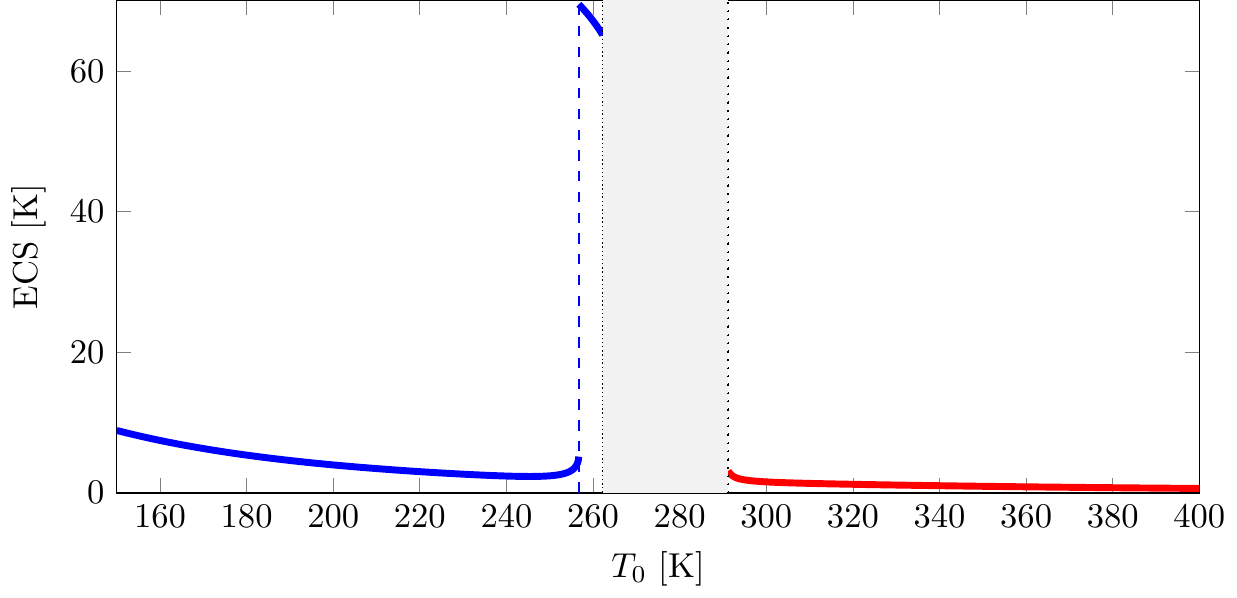}
         \caption{Equilibrium Climate Sensitivity vs initial equilibrium temperature $T_0$.}
    \end{subfigure}
   
    \caption{(a) Bifurcation diagram for \eqref{eq:0DGEBM} with parameters A from Table~\ref{tab:fsGEBM}. The bifurcation parameter $\mu$ represents radiative forcing due to atmospheric \COO. Solid lines correspond to stable equilibria, and dashed lines to unstable equilibria. There are two different branches of stable equilibria: one that corresponds to a cold climate (blue) and one that corresponds to a warm climate (red). (b) Equilibrium `two-point' response for different forcing levels corresponding to $2^\gamma \times$\COO\, starting from an initial state $T_0 = 293K$ corresponding to equilibrium temperature before perturbation. The red part of the figure correspond to end states on the warm branch; the blue part to end states on the cold branch. The dashed line indicates the location of a tipping point. (c) ECS (i.e. equilibrium two-point response to \COO\ doubling) as function of the initial temperature $T_0$. Blue lines indicate starting points on the cold branch; red lines indicate starting point on the warm branch; the grey region corresponds to unfeasible initial temperatures (i.e., they lie on the unstable branch in (a)). The large peak in the blue line corresponds to tipping from the cold branch to the warm branch; the location of this tipping point is indicated with a dashed line.}
    \label{fig:0DGEBM}
\end{figure}

It is well known that in the case of no internal variability, equations  (\ref{eq:0DGEBM}) can be bistable~\cite{budyko, ghil1976climate}. Due to the functional forms of temperature dependent albedo and emissivity, the model (\ref{eq:0DGEBM}) can have one, two or three stable equilibria depending on the parameter values. This is organized by a fifth order ``butterfly'' singularity \cite{montaldi2021}: see Appendix~\ref{app:butterfly} for a verification and in-depth analysis of the bifurcation structure of this model. Nonetheless, for the parameter values given in Table~\ref{tab:fsGEBM}, the model is bistable for a certain range of values of the parameter $\mu$: in this bistable region, the model supports a stable cold ``icehouse'' and a warm ``hothouse'' climate state (see Figure~\ref{fig:0DGEBM}).

\begin{table}
$$
\begin{array}{l|cccc|l}
& \mbox{A} & \mbox{B}& \mbox{C}& \mbox{D}&  \mbox{units} \\
\hline
C & 5\times 10^8 &\cdots & \cdots & \cdots &  Jm^{-2}K^{-1}
\\
Q_0 & 341.3 &\cdots & \cdots & \cdots &Wm^{-2}
\\
\sigma & 5.67 \times 10^{-8} & \cdots & \cdots & \cdots &Wm^{-2}K^{-4}
\\
\alpha_1 & 0.7 &\cdots & \cdots & \cdots &
\\
\alpha_2 & 0.289 &\cdots & \cdots &\cdots &
\\
T_{\alpha} & 274.5 &\cdots & \cdots &\cdots & K
\\
K_{\alpha} & 0.1 & \cdots & \cdots & \cdots &K^{-1}
\\
\varepsilon_1 & 0.5& \cdots & \cdots & \cdots &
\\
\varepsilon_2 & 0.41& \cdots & \cdots & \cdots &
\\
T_{\varepsilon} & 288 & \cdots & \cdots &\cdots & K
\\
K_{\varepsilon} & 0.5 &\cdots & \cdots & 0.1 & K^{-1}
\\
A_0 & 5.35 & \cdots & \cdots &\cdots & Wm^{-2} 
\\
\tau_{\alpha} & 0 & 0 & 5\times 10^{9} & 5\times 10^{9} & s
\\
\tau_{NV} & 0 & 6\times 10^7 & 6\times 10^7 & 6\times 10^7 & s
\\
\nu_{NV} & 0 & 5 & 2\times 10^{-2} & 2\times 10^{-2} & Wm^{-2}
\end{array}
$$
\caption{Values for the Global Energy Balance Model (GEBM) (\ref{eq:0DGEBM}) with dynamic albedo (\ref{eq:fsalpha}) and chaotic variability (\ref{eq:chaos}) used in the numerical simulations. We take the standard choice for the Lorenz parameters: $\sigma = 10$, $\rho = 28$ and $\beta = 8/3$. For the simulations, time $t$ is rescaled to years. The equilibrium albedo is given by (\ref{eq:alpha0}) and (equilibrium) emissivity by (\ref{eq:varepsilon}). The forcing $\mu$ is given by (\ref{eq:mu}) and $\nu_{NV}$ represents the amplitude of a chaotic forcing via (\ref{eq:NV}). The use of ``$\cdots$'' indicates that values of column $\mbox{A}$ are also used in this case.}
\label{tab:fsGEBM}
\end{table}

Note that the ECS of both type of states (for the same \COO-level) differs between branches as albedo and emissivity are different between branches. However, the ECS within a branch is also not constant: Figure~\ref{fig:0DGEBM}(b) shows variation between initial points $y_0$ that lie on the same branch (intra-branch differences). In the climate literature, these variations are not well-quantified, mainly because they depend on a multitude of physical feedback processes, which are difficult to observe and model numerically in full \cite{Sherwood.2020,Heydt.2016}. Still, it is good to keep in mind that observed or estimated ECS might vary as the (initial) climate state changes.

\subsection{Nonlinear response: slow and/or late tipping and ECS}

\begin{figure}
	\centering
	\begin{subfigure}[t]{\textwidth}
		\includegraphics[width=\textwidth]{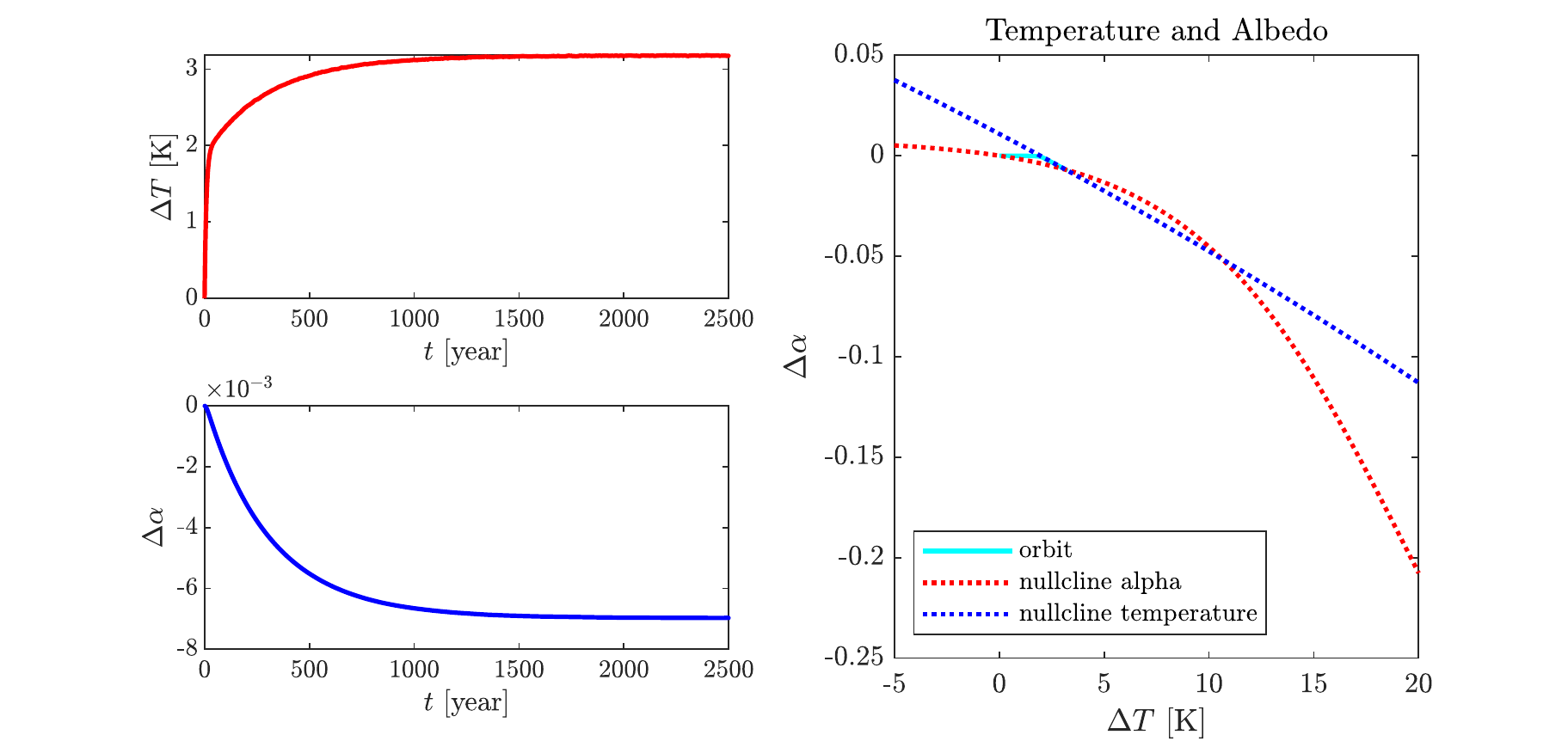}
		\caption{time series and phase diagram}
	\end{subfigure}\\
	
	\begin{subfigure}[t]{0.45\textwidth}
		\includegraphics[width=\textwidth]{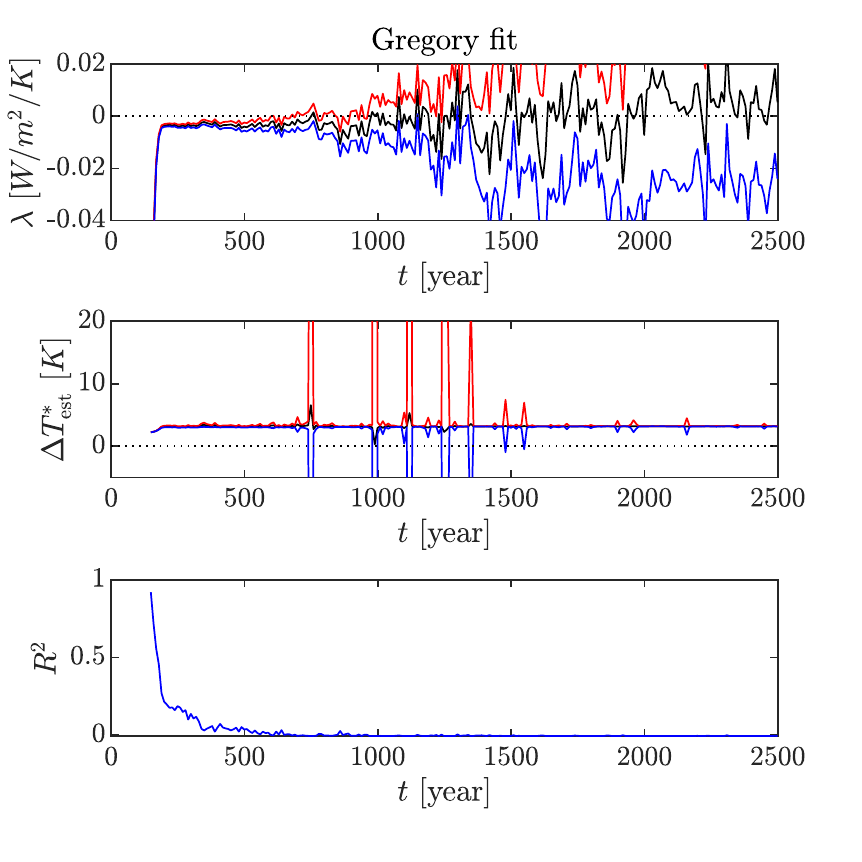}
		\caption{Gregory Fit}
	\end{subfigure}~
	\begin{subfigure}[t]{0.45\textwidth}
		\includegraphics[width=\textwidth]{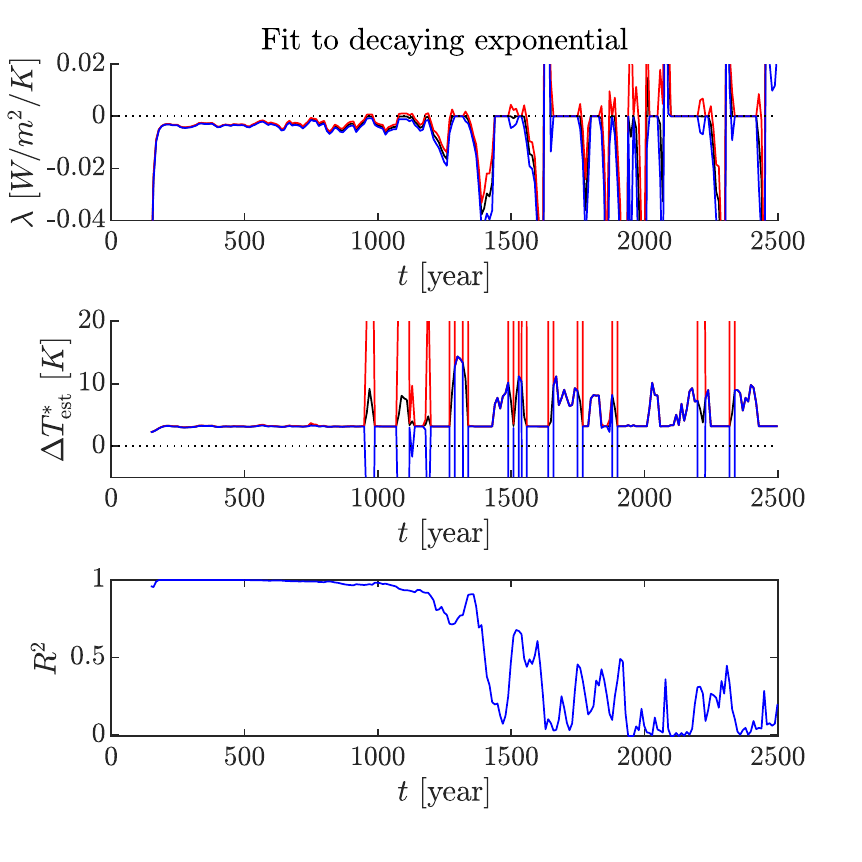}
		\caption{Exponential Fit}
	\end{subfigure}
	
	\caption{Warming in the fast-slow energy balance model (parameters C in Table~\ref{tab:fsGEBM}) subjected to an abrupt2xCO2 forcing. The initial forcing $\mu_0$ is chosen such that there is an initial equilibrium is $T_0 = 255K$. (a) Time series for $\Delta T$ and $\Delta \alpha$ as well as a the trajectory through (projected) phase space (cyan). The red dotted curve in the right panel denotes the nullcline on which $\frac{d\alpha}{dt} = 0$ and the blue dotted curve denotes the nullcline on which $\frac{dT}{dt} = 0$, which also acts as a slow manifold. 
	(b) Gregory fits on time windows of $150$ years, showing the thus estimated feedback parameter $\lambda$ and expected equilibrium warming $\Delta T_\mathrm{est}^*$ over time, together with standard errors and $R^2$ statistic from the linear regression. (c) Fit to a decaying exponential on time windows of $150$ years, showing the estimated feedback parameter $\lambda$ and expected equilibrium warming $\Delta T_\mathrm{est}^*$ over time, similarly showing upper/lower standard error estimates and $R^2$ from the nonlinear regression.
    }
	\label{fig:fastslowGEBM2xCO2}
\end{figure}

As discussed in section~\ref{sec:EqResponse}, if the transient relaxation dynamics of a climate model is approximated well by a linear system this can be used to estimate ECS. This thus works for nonlinear systems with small enough forcings. For example Figure~\ref{fig:fastslowGEBM2xCO2}(a) shows a simulation of (\ref{eq:0DGEBM}) with parameters C in Table~\ref{tab:fsGEBM} subjected to an abrupt2xCO2 forcing. The initial forcing $\mu_0$ is chosen such that there is an equilibrium at $T_0 = 255K$. There is a clear two-stage exponential decay to equilibrium with $\Delta T^*\approx 3.1$K: the right panel of Figure~\ref{fig:fastslowGEBM2xCO2}(a) shows that there is a nearby equilibrium attractor. 

Figure~\ref{fig:fastslowGEBM2xCO2}(b) shows estimates using the Gregory method on rolling windows of $150$ years. Within a time window, we regress the time series of $\Delta T$ and $\Delta N := C \frac{d\Delta T}{dt}$ to the linear model $\Delta N = f + \lambda \Delta T$, which gives estimates for the forcing $f$ and the dominant feedback parameter $\lambda$. The regression is performed using the MATLAB fit to linear model {\tt fitlm}; standard errors for best fit are shown, and the bottom panel shows the adjusted $R^2$-statistic for this window, where $R^2=1$ implies all variance in the signal is described by the model within the window ending at that time-point.  Equilibrium warming is derived from these fits by extrapolation of the linear model, giving $\Delta T^*_\mathrm{est} = - f / \lambda$.  Note that the initial 150 year fit is already good. Indeed one can see decreasing signal-to-noise ratio and $R^2$ for fits taken later in the time series, as noise dominates the dynamics of the state this late in the simulation. 

A second equilibrium estimation protocol is shown in Figure~\ref{fig:fastslowGEBM2xCO2}(c) in which blocks of 150 years are fitted to a decaying exponential function $T(t) = T_{\infty}+b e^{\lambda t}$ using the MATLAB fit to nonlinear model {\tt fitnlm}. This gives an estimate for $T_{\infty}$, $b$ and $\lambda$ with standard errors and is a direct approximation of a linear response to a Heaviside input; we show $\lambda$ and $\Delta T_{est}^*=T_{\infty}-T_0$. Observe that, similarly to the Gregory fits, these fits also become degenerate for later time frames. 

\begin{figure}
    \centering
        \begin{subfigure}[t]{\textwidth}
            \includegraphics[width=\textwidth]{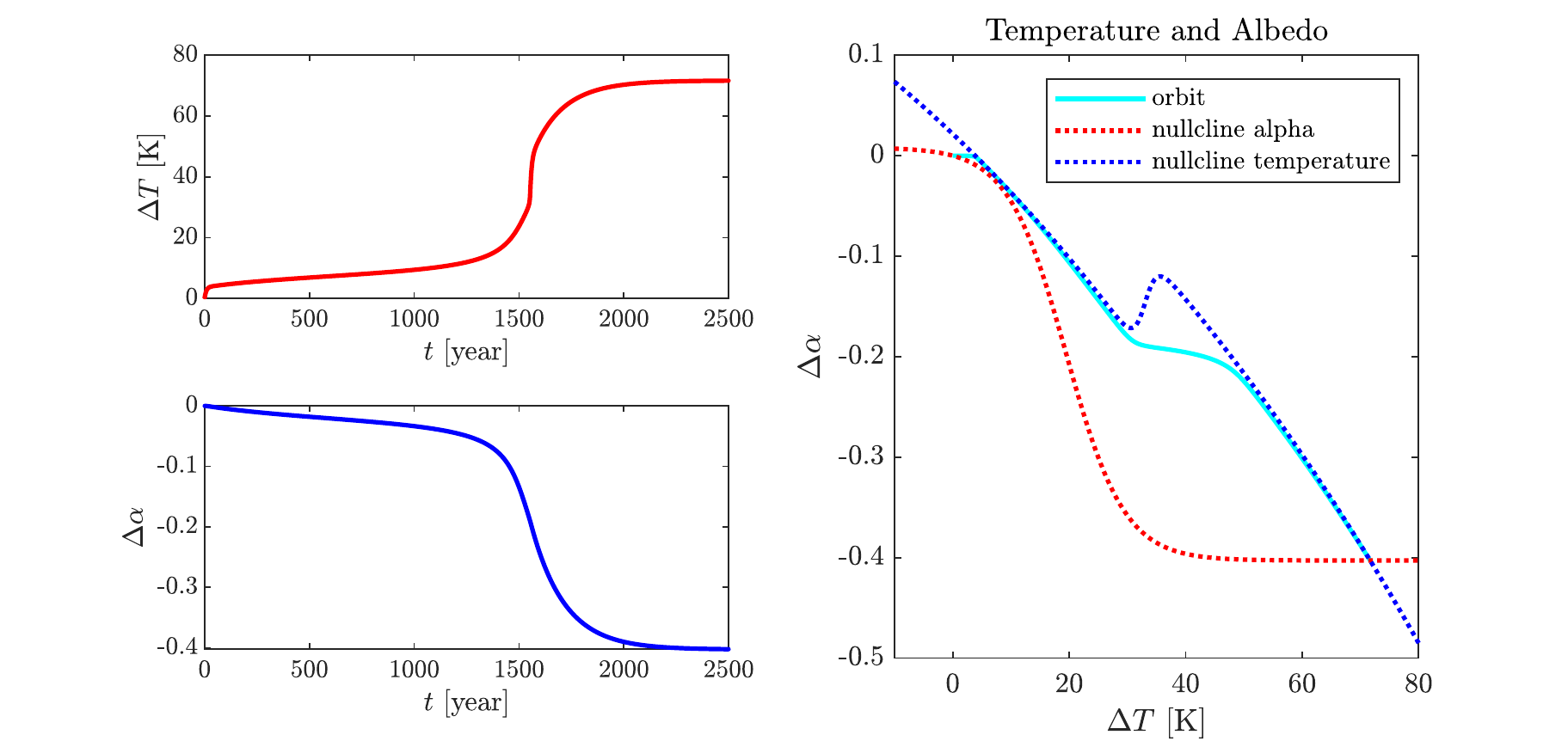}
            \caption{time series and phase diagram}
        \end{subfigure}\\
        
        \begin{subfigure}[t]{0.45\textwidth}
            \includegraphics[width=\textwidth]{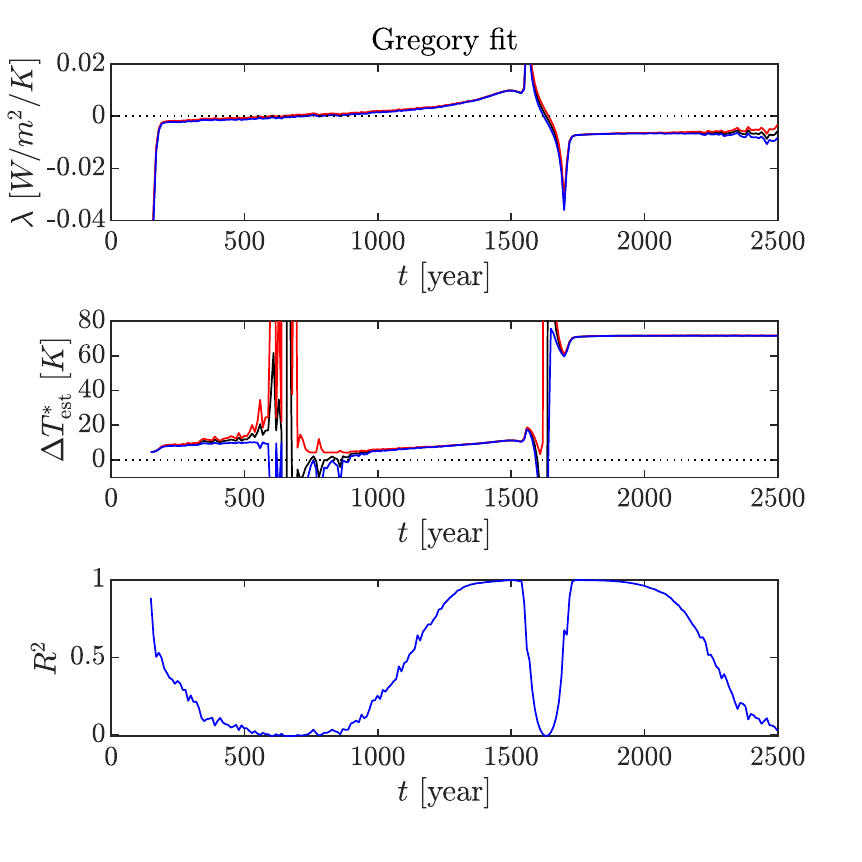}
            \caption{Gregory Fit}
        \end{subfigure}~
        \begin{subfigure}[t]{0.45\textwidth}
            \includegraphics[width=\textwidth]{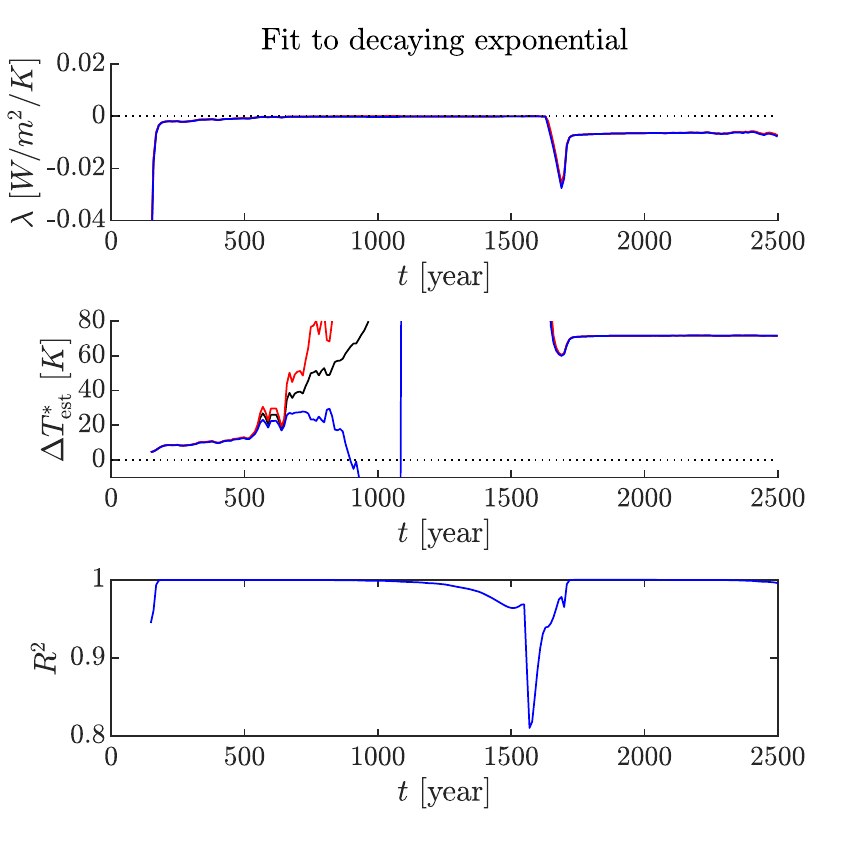}
            \caption{Exponential Fit}
        \end{subfigure}

    \caption{Warming in the fast-slow energy balance model (parameters C in Table~\ref{tab:fsGEBM}) subjected to an abrupt4xCO2 forcing. The initial equilibrium is $T_0 = 255K$. Here, a late tipping event happens as the dynamics drive the system over a fold point of the slow manifold. (a) Time series for $\Delta T$ and $\Delta \alpha$ as well as a the trajectory through phase space (cyan). The red dotted curve in the right panel denotes the nullcline on which $\frac{d\alpha}{dt} = 0$ and the blue dotted curve denotes the nullcline on which $\frac{dT}{dt} = 0$, which also acts as a slow manifold. 
    Fits (b,c) as in Figure~\ref{fig:fastslowGEBM2xCO2}.
    }
    \label{fig:fastslowGEBM}
\end{figure}

To contrast with Figure~\ref{fig:fastslowGEBM2xCO2}, Figure~\ref{fig:fastslowGEBM} shows the case for an abrupt 4xCO2 forcing but otherwise identical parameters and initial condition, in which the transient dynamics are not approximated well by a linear system, although a long transient period (due to the crossing of a slow tipping point) conceals the nonlinear dynamics. Figure~\ref{fig:fastslowGEBM}(a) shows that the run seems to rapidly approach an equilibrium, but warming then continues slowly as albedo slowly decreases. Then, around $t=1500$ years, there is a surprising and rapid ``late tipping'' followed by a relaxation to the final equilibrium. From the fits in Figure~\ref{fig:fastslowGEBM}(b) and (c), approximately linear behaviour can be seen at first; however, we are near (but beyond) a fold bifurcation on the stable part of the slow manifold where the blue and red nullclines become tangent (i.e. a slow tipping point), and for this forcing the nullclines are barely detached. As the state passes this point (sometimes called a ghost attractor), the dynamics on the slow manifold speed up before tipping over a fold in the slow manifold, causing a rapid late tipping event to another stable branch of this slow manifold. 

Figure~\ref{fig:fastslowGEBM}(b) shows estimates using a Gregory fit. It can be seen that a fast decay is picked up initially, and slower decay dominates from about $t = 250$ years. At around $t = 500$ years, the fitted value for $\lambda$ passes through zero, suggesting a linearly unstable climate, and the estimated warming becomes unreliable. Only after the late tipping event, from $t = 1750$ years onwards, the fits make sense again, with negative $\lambda$ and sensible warming estimates corresponding to the actual equilibrium warming of the simulation. Similarly, for the exponential fit shown in (c), $\lambda \approx -0.05$ corresponding to the initial fast decay but this quickly decays to pick up the slow decay with $\lambda \approx -0.001$ by about $t=250$. The fit remains good up to $t\approx 500$ years but after this the estimated errors on $\Delta T_{est}$ increase rapidly as the fit attempts to fit a decaying exponential to something that is actually growing slowly but exponentially. At $t\approx 1500$ years the system passes through the late rapid tipping before settling to a fit to $\Delta T_{est}^*\approx 72$.

Clearly, in both of these fitting approaches the true equilibrium warming is not estimated accurately at all until after the late tipping event when the system is again approximately linear. From the fits up to about $t = 500$ years there are no obvious hints that anticipate this late tipping and the fit results seem to indicate convergence to a noisy equilibrium state (hence for example the low $R^2$ score for the Gregory method as it is mostly noise at this point). Only after $t = 500$ years there start to be some signs of the passing of a slow tipping point ($\lambda > 0$ in the Gregory method and large uncertainties in the exponential fit method) in this example, as the almost-equilibrium (ghost attractor) on the slow manifold is passed around this time.

Comparing Figures~\ref{fig:fastslowGEBM} and \ref{fig:fastslowGEBM2xCO2}, we see very similar fits and estimates up to $t = 500$ years, further indicating the difficulty of distinguishing scenarios with and without late tipping. Moreover, the perturbation that exceeds the threshold shown in Figure~\ref{fig:tradeoff5}(b) lies somewhere between 2xCO2 and 4xCO2 for this model and parameters.

\begin{figure}
    \centering
        \begin{subfigure}[t]{\textwidth}
            \includegraphics[width=\textwidth]{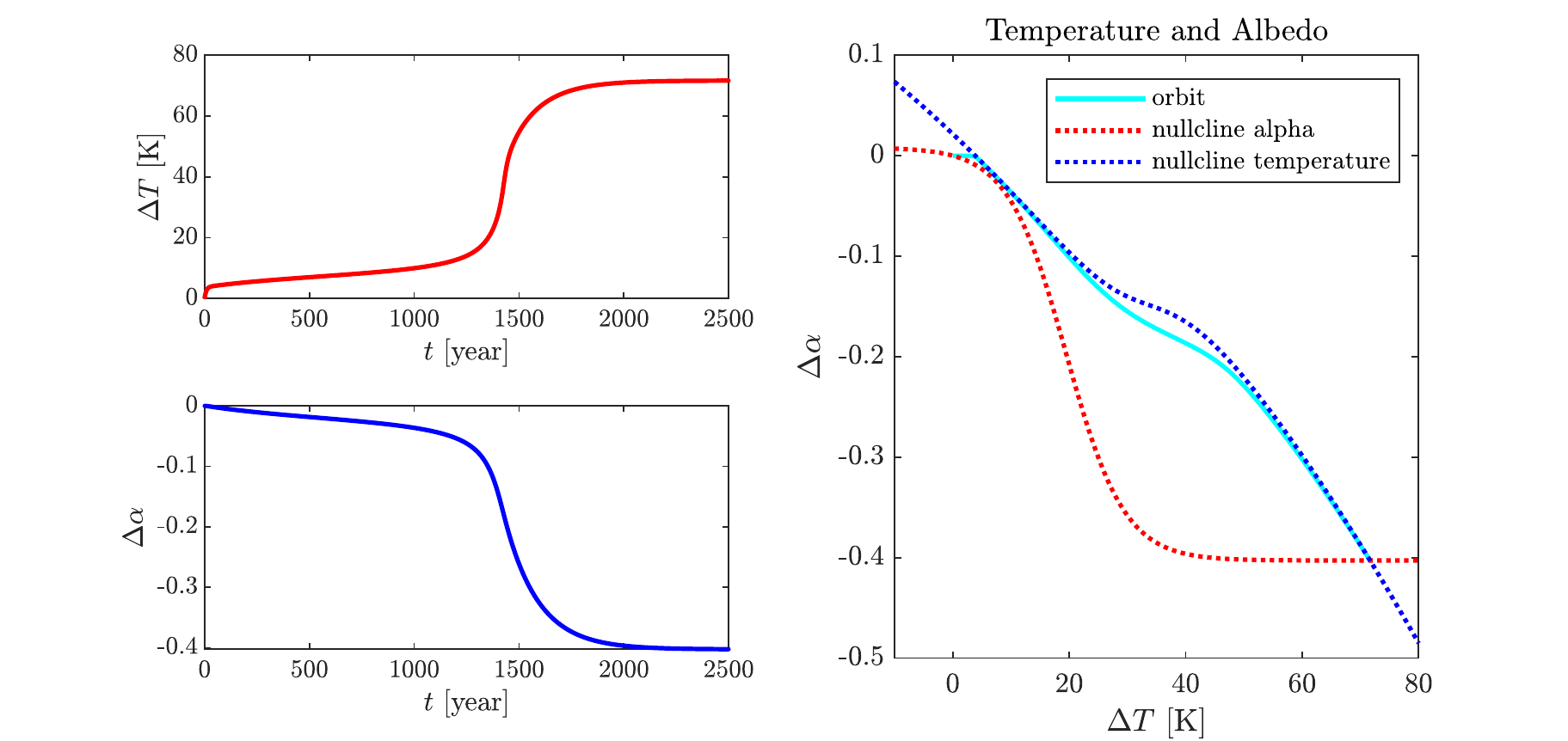}
            \caption{time series and phase diagram}
        \end{subfigure}\\
        
        \begin{subfigure}[t]{0.45\textwidth}
            \includegraphics[width=\textwidth]{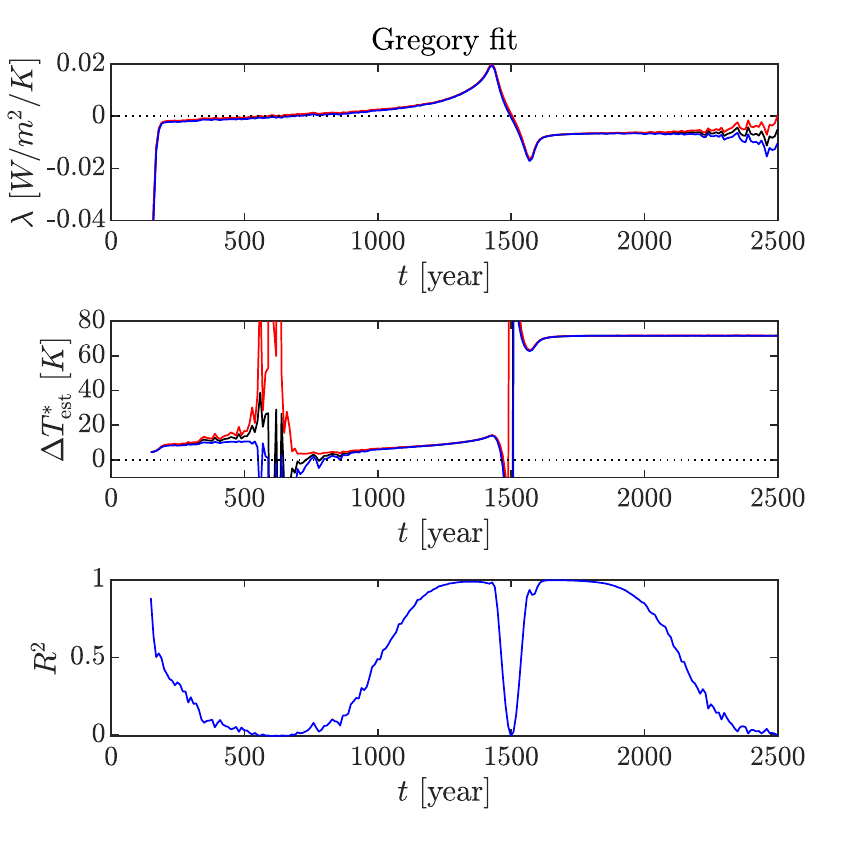}
            \caption{Gregory Fit}
        \end{subfigure}~
        \begin{subfigure}[t]{0.45\textwidth}
            \includegraphics[width=\textwidth]{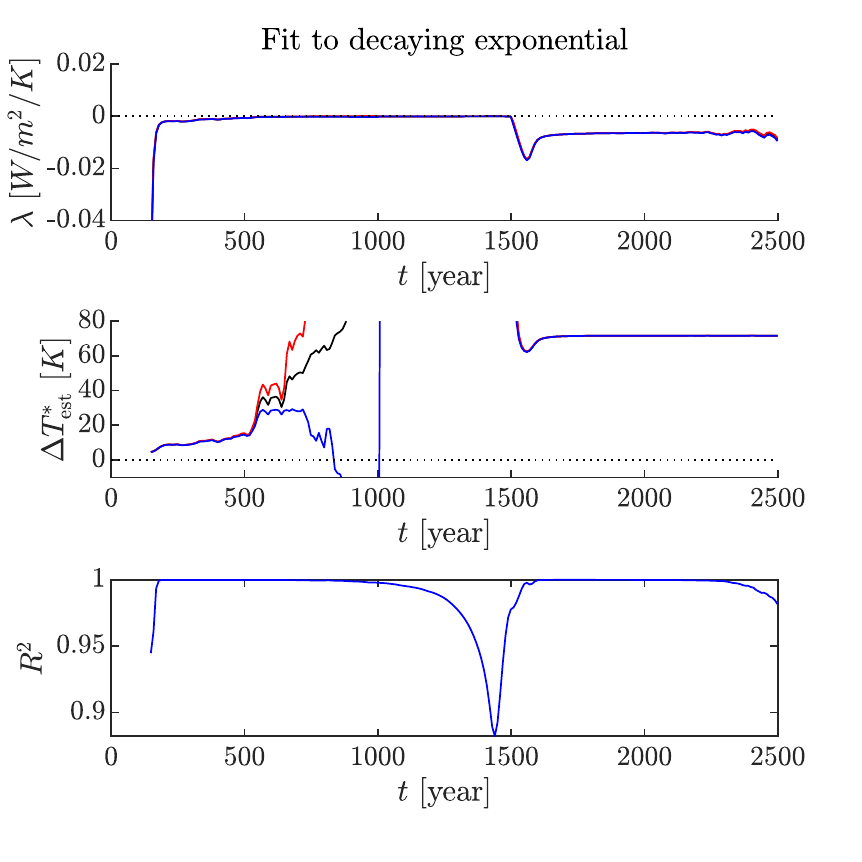}
            \caption{Exponential Fit}
        \end{subfigure}

    \caption{(a) Warming in the fast-slow energy balance model (parameters D in Table~\ref{tab:fsGEBM}) subjected to an abrupt4xCO2 forcing. The initial equilibrium is $T_0 = 255K$. The left panel shows time series for $\Delta T$ and $\Delta \alpha$. The right panel shows the trajectory through phase space (cyan). The red dotted curve in the right panel denotes the nullcline on which $\frac{d\alpha}{dt} = 0$ and the blue dotted curve denotes the nullcline on which $\frac{dT}{dt} = 0$, which also acts as a slow manifold. Note that for the parameters D there is no longer a late tipping in $T$ of the speed as seen in Figure~\ref{fig:fastslowGEBM}, nonetheless there is a moderately rapid increase in $T$ around 1500 years. Fits in (b,c) as for Figure~\ref{fig:fastslowGEBM2xCO2}.
    }
    \label{fig:fastslowGEBMnolate}
\end{figure}

Figure~\ref{fig:fastslowGEBMnolate} shows an analogous simulation of (\ref{eq:0DGEBM}) under abrupt4xCO2 forcing but parameters D of Table~\ref{tab:fsGEBM}. Again, the initial forcing $\mu_0$ is such that there is an initial equilibrium at $T_0=255K$. For these parameters there is no fold in the critical manifold meaning that there is not a rapid late tipping (in the bifurcation sense).However, similarly to Figure~\ref{fig:fastslowGEBM}, the initial (linear) warming is not representative of the equilibrium warming and the transient means one can only see evidence of the final state after $t=1500$ years. 

This indicates that even in the absence of (late) tipping points, an initial good fit cannot exclude a later rapid warming phase in systems that have dynamics on multiple time scales. For all three simulations presented in this section, extrapolations from fits to the initial few hundred years look very similar, although their long-term behaviour is very different, again highlighting that extrapolations may only be accurate after long transients that bring the system into a linear regime.

\subsection{Ensemble variability and ECS}

When estimating ECS in models with internal variability, one of the ingredients is the precise choice of the initial conditions $y_0$. In Section~\ref{sec:EqResponse}\ref{sec:background_state}, we already discussed that the background climate state (i.e., the initial attractor $A_0$) influences the transient and equilibrium response to forcings. However, also the precise initial state $y_0$ on the initial attractor will impact the observed transient dynamics and can potentially also change the final equilibrium state. We illustrate such situations in this subsection.

\begin{figure}
    \centering
    \begin{subfigure}[t]{0.4\textwidth}
        \centering
        \includegraphics[width=\textwidth]{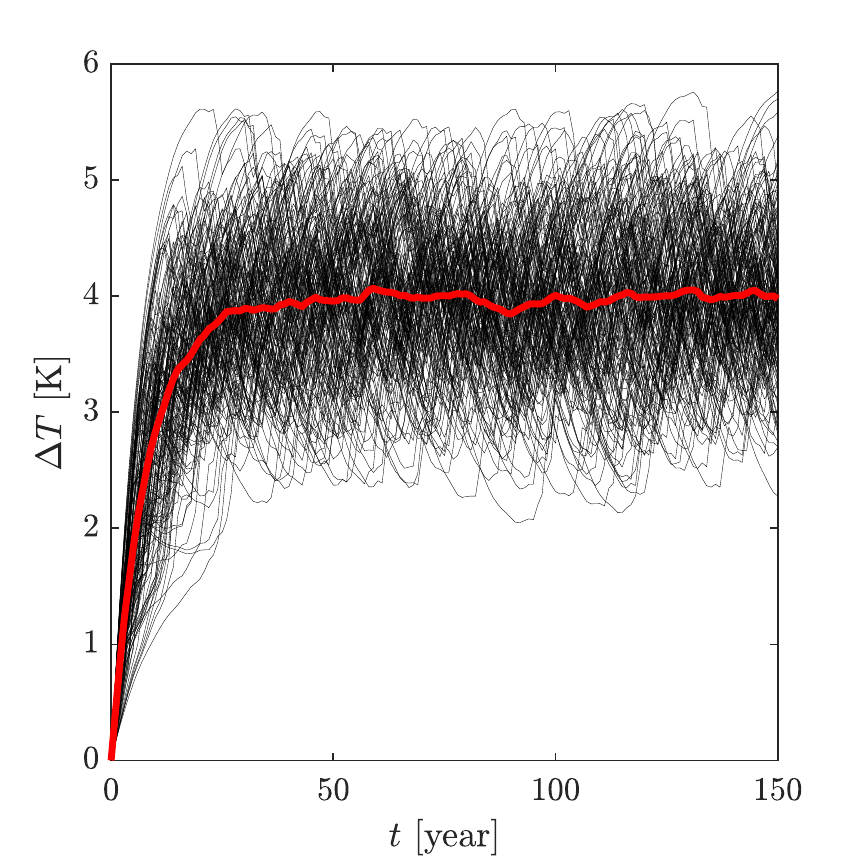}
        \caption{time series}
    \end{subfigure}
    \begin{subfigure}[t]{0.4\textwidth}
        \centering
        \includegraphics[width=\textwidth]{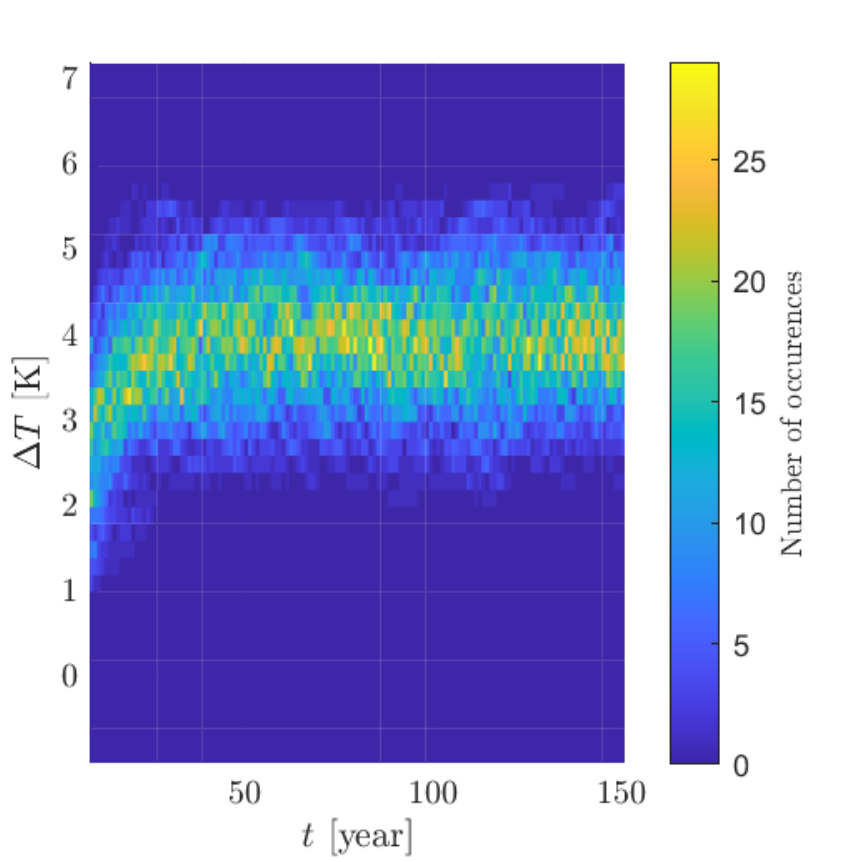}
        \caption{heat map}
    \end{subfigure}
     \begin{subfigure}[t]{0.4\textwidth}
        \centering
        \includegraphics[width=\textwidth]{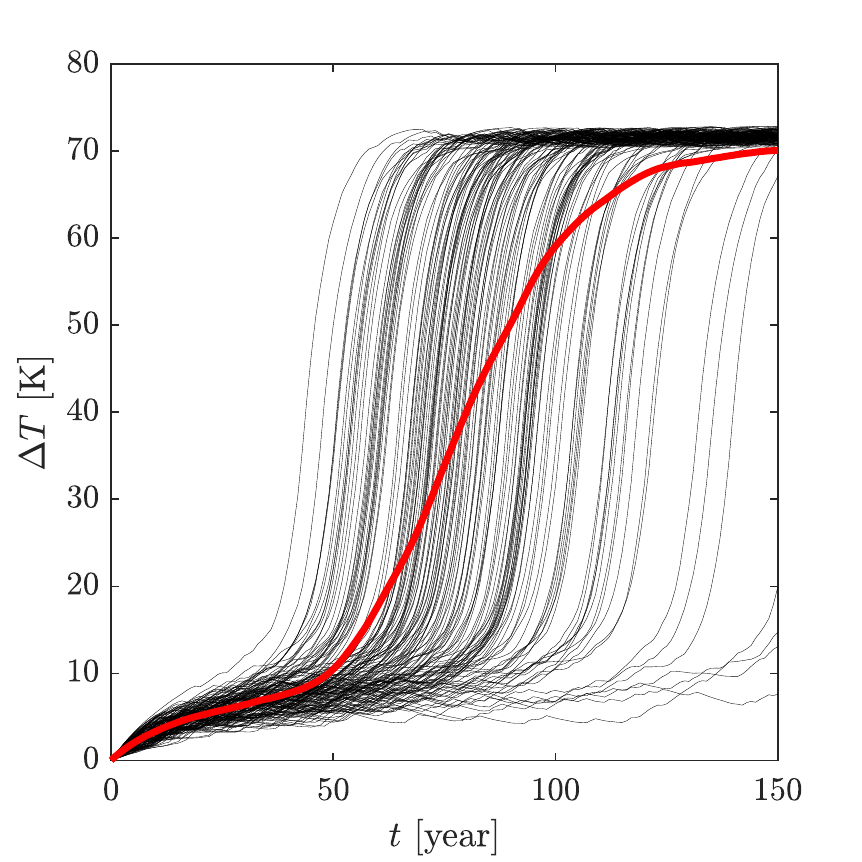}
        \caption{time series}
    \end{subfigure}
    \begin{subfigure}[t]{0.4\textwidth}
        \centering
        \includegraphics[width=\textwidth]{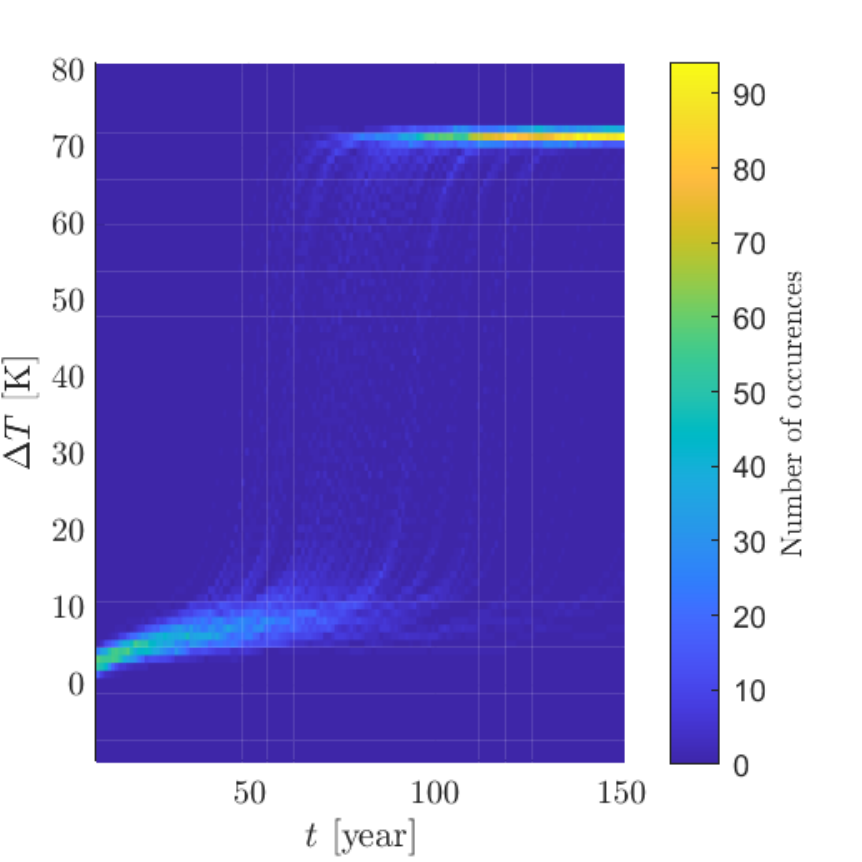}
        \caption{heat map}
    \end{subfigure}

    \caption{Results of a $150$ ensemble run of abrupt4xCO2 experiments for the energy balance model with chaotic forcing and instantaneous albedo relaxation (parameters B of Table~\ref{tab:fsGEBM} with initial temperature (a,b) $T_0 = 293K$ (a warm climate) and (c,d) $T_0=255K$ (a cold climate). Initial conditions for the Lorenz part of the model are randomly chosen for each ensemble member separately. (a,c) Time series of the warming $\Delta T$ over time for a random set of $100$ of the ensemble members (black) and the ensemble average (red). (b,d) Heat maps indicating the number of times a certain warming has been observed per time step (note that temperature bins are differently sized between (b) and (d)). }
    \label{fig:chaoticGEBM}
\end{figure}

\begin{figure}
    \centering
    \begin{subfigure}[t]{0.49\textwidth}
        \centering
        \includegraphics[width=\textwidth]{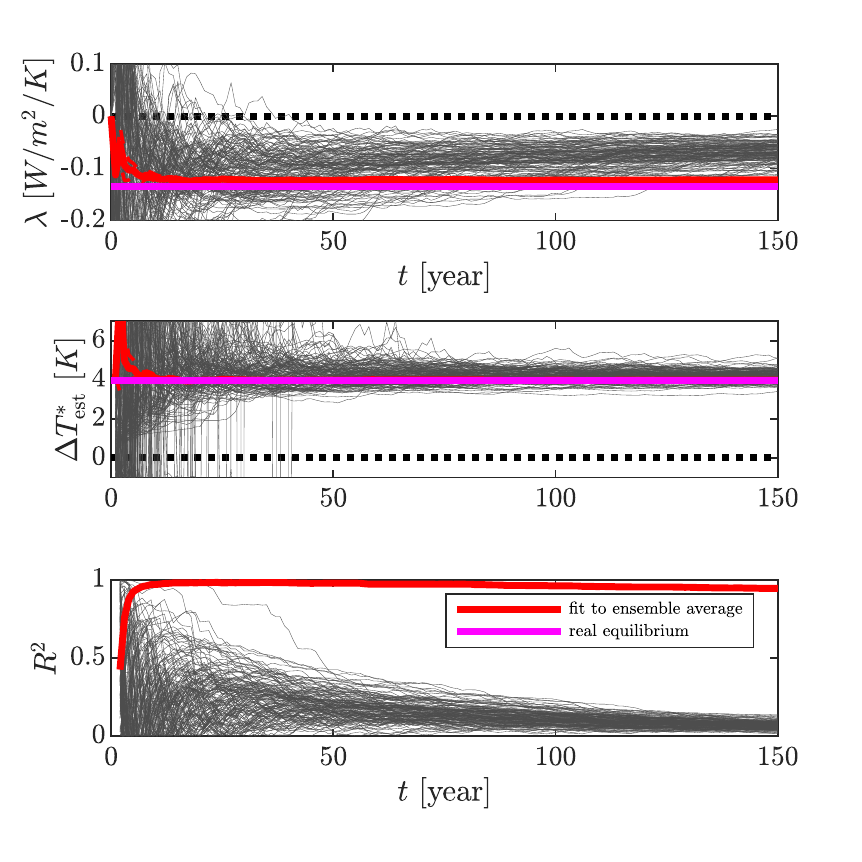}
        \caption{}
    \end{subfigure}
    \begin{subfigure}[t]{0.49\textwidth}
        \centering
        \includegraphics[width=\textwidth]{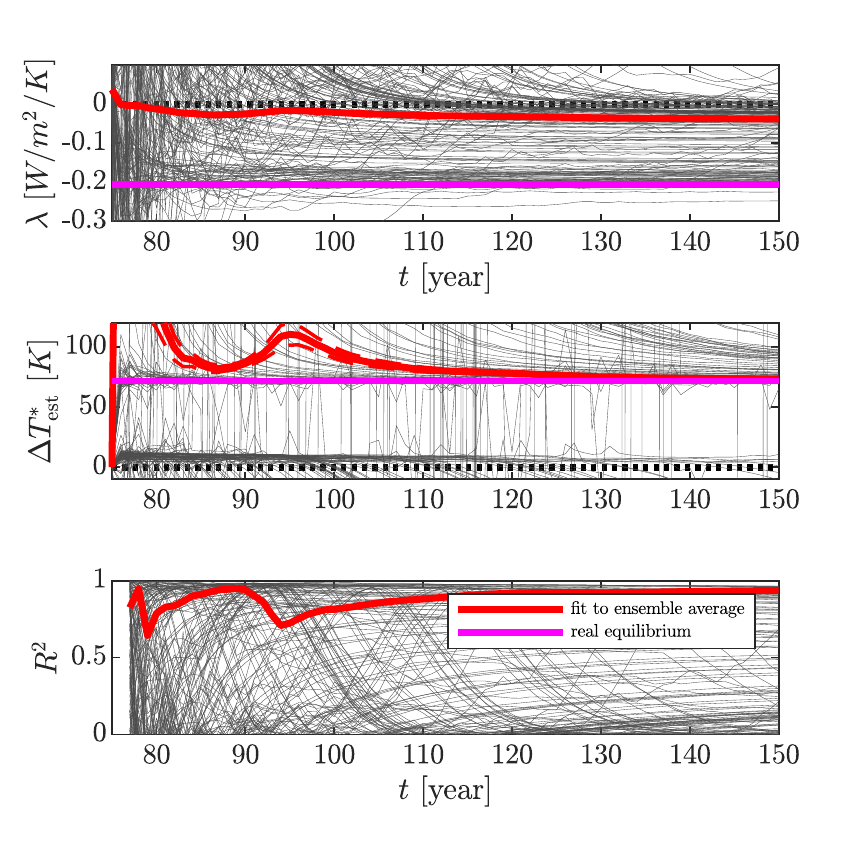}
        \caption{}
    \end{subfigure}
    
    \caption{
    (a) Gregory fits for ensemble data from Figure~\ref{fig:chaoticGEBM}(a,b) up to year $t$, showing the feedback parameter $\lambda$, the estimated equilibrium warming $\Delta T_\mathrm{est}^*$, and the $R^2$ statistic. (b) similar for Figure~\ref{fig:chaoticGEBM}(c,d). Grey lines indicate results on individual ensemble members; red lines indicate results of regression on the ensemble mean;  magenta lines indicate the theoretical real values. The solid lines indicate the expected values and the dashed lines the standard errors (almost imperceptible as standard errors are typically very low). The black dotted line denotes the zero-line.
    }
    \label{fig:chaoticGEBM_gregory}
\end{figure}

Figure~\ref{fig:chaoticGEBM} (a,b) show an abrupt4xCO2 experiment for an ensemble of different initial states on the same initial attractor (a warm climate state), for a simulation of~\eqref{eq:0DGEBM} with parameters B of Table~\ref{tab:fsGEBM} -- note the presence of chaotic variability. There is potential variation in the warming of the different ensemble members during the transient, which stems from different realisations of the natural variability, corresponding to the different initial states.  Gregory fits over a time window starting at time $0$ up to time $t$ are shown in Figure~\ref{fig:chaoticGEBM_gregory}(a). The associated regression to individual ensemble members (black) are poor, but the regression to the ensemble average (red) is much better as the noise (internal variability) is averaged out.

Another example is given in Figure~\ref{fig:chaoticGEBM}(c,d) for a different initial attractor (a cold climate state), where natural variability pushes the state over a tipping point at different times during the simulation of each ensemble member. The simulations initially suggest relaxation towards a state close to the original colder state, but later they consistently exhibit tipping to a different (and much warmer state). In this example the colder state is almost at equilibrium. As long as the natural variation in forcing is small enough, the system remains close to the colder state. For larger fluctuations the system tips into the warmer state. 
Figure~\ref{fig:chaoticGEBM_gregory}(b) shows that even the ensemble average is not adequate to estimate ECS in this case; accurate estimates can only be made if the model has been run until (almost) all individual ensemble members have tipped. Nevertheless, the ensemble averaged response is still much better than the other approaches, because data from tipped and non-tipped ensemble members leads otherwise to very unreliable bimodal estimates with high variance.

\begin{figure}
    \centering
    \begin{subfigure}[t]{0.4\textwidth}
        \centering
        \includegraphics[width=\textwidth]{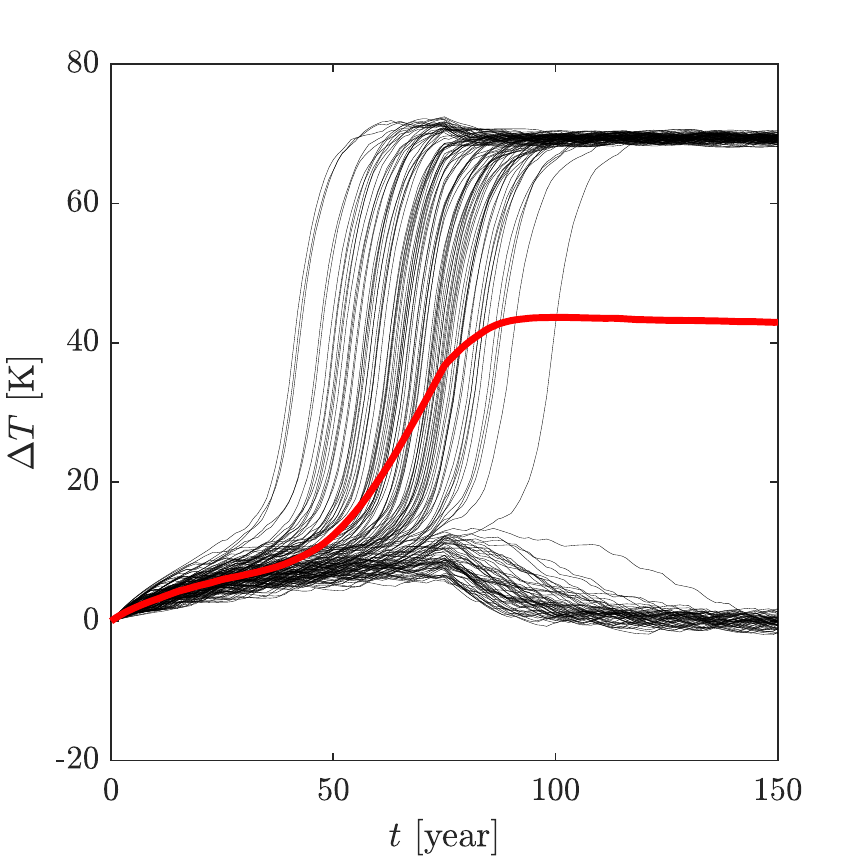}
        \caption{time series}
    \end{subfigure}
    \begin{subfigure}[t]{0.4\textwidth}
        \centering
        \includegraphics[width=\textwidth]{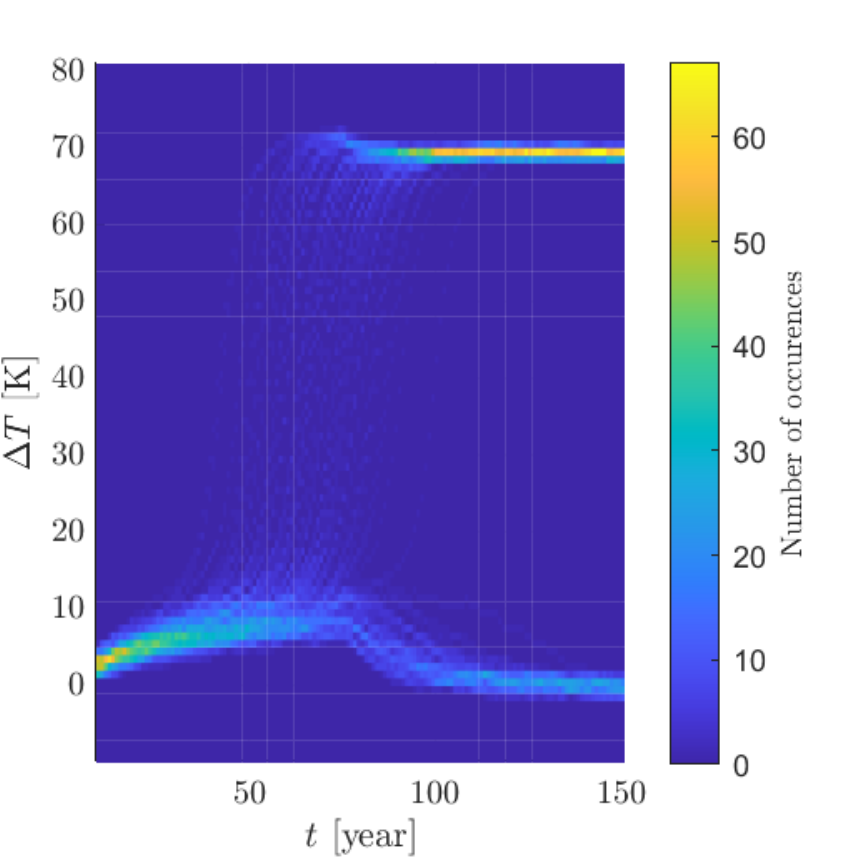}
        \caption{heat map}
    \end{subfigure}
    
    \caption{Results of a $150$ ensemble run of an experiment for the energy balance model with chaotic forcing and instantaneous albedo relaxation (parameters in column B in table~\ref{tab:fsGEBM}) in which there is a temporary \COO\ quadrupling forcing for $75$ years after which the initial \COO-levels are restored. The initial temperature is $T_0 = 255K$ (a cold climate).(a) Time series of the warming $\Delta T$ over time for a random set of $100$ of the ensemble members (black) and the ensemble average (red). (b) Heat map indicating the number of times a certain warming has been observed per time step (temperature bins of size $1K$ haven been used). 
	}
    \label{fig:chaoticGEBM_partial_tipping}
\end{figure}

Even worse, for non-constant forcing the equilibrium response may depend more drastically on the precise initial state $y_0$; some part of the initial attractor $A_0$ can be attracted to a final attractor $A_1$, while the rest is attracted to a different final attractor $\tilde{A}_1$. This effect has been called a \emph{partial tipping} of the attractor and is studied abstractly in \cite{alkhayuon2018rate,ashwin2020extreme,ashwin2021physical}. Because of the relative simplicity of the chaotic GEBM~\eqref{eq:0DGEBM}, we cannot show this behaviour for constant forcing, but we can illustrate this phenomenon by forcing the model temporarily with an abrupt4xCO2 forcing, after which the initial \COO-levels are restored at time $t = 75$ years.  Figure~\ref{fig:chaoticGEBM_partial_tipping} shows the results of this experiment. One can clearly see that some ensemble members experience tipping but others do not. In this situation (details not shown), partial tipping means that none of the ECS estimation techniques will paint a full picture. The ensemble-average does contain some information on the number of tipping and non-tipped states but we suggest more meaningful estimates would need to be made for the attractors separately, first by categorising each individual ensemble member as tipped or not, and using estimation techniques on these categories separately.

\subsection{Evidence of late tipping within GCM runs}
\label{sec:longrunmip}

For GCM runs with conditions corresponding to the relatively stable conditions of the Holocene pre-industrial climate, the accepted wisdom is that we do not expect to find any major global tipping effects as extreme as the ice-house to hothouse transitions explored above. Nonetheless there are hints that we may be close to regional tipping points such as changes in the Atlantic Meridional Overturning Circulation (AMOC) or West Antarctic icesheet collapse, and some emissions scenarios are likely to take us over these tipping points. Crossings of these regional tipping points can result in a global signal, such as changes in the AMOC leading to global climatic changes~\cite{stouffer2006investigating, jackson2015global}. Further, as emission reduction scenarios may take us over tipping points only temporarily \cite{Ritchie.2021}, also the possibility of a partial tipping of an attractor may be very relevant to study in GCMs.

Initial conditions for GCM runs are notoriously difficult to set -- they are typically taken as the end of a spin-up simulation, or as a state at some time during a control experiment (in both of which atmospheric \COO\ is kept fixed at the starting levels). In ensemble runs, variation of initial states on the initial attractor are sometimes explored either by sightly perturbing an initial state (called `micro-perturbations'), or by taking several states of a control run, typically separated by a few months up to a few years, depending on the time scale of the internal variability that is being considered (called `macro-perturbations')~\cite{deser2020insights, aragon2016mip}. Nonetheless, even after substantial spin-up there may be continued variability that can cause extrapolations such as Effective Climate Sensitivity to continue varying over centennial timescales \cite{senior2000time}. For example, \cite{manabe1993century,manabe1994multiple} find multi-century changes in an atmosphere-ocean GCM, mostly to do with the strength of the AMOC, depending on the magnitude of the \COO\ perturbation.

\begin{figure}
	\centering
	\begin{subfigure}[t]{0.45\textwidth}
		\includegraphics[width=\textwidth]{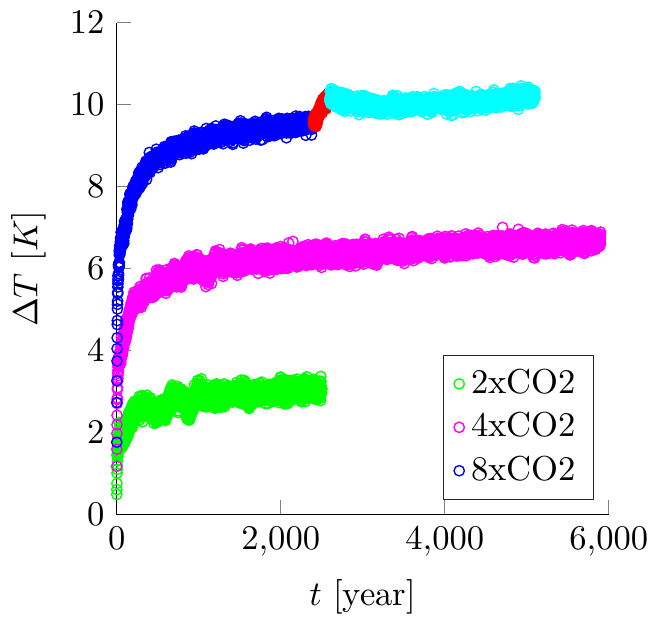}
		\caption{time series}
	\end{subfigure}
	\begin{subfigure}[t]{0.45\textwidth}
		\includegraphics[width=\textwidth]{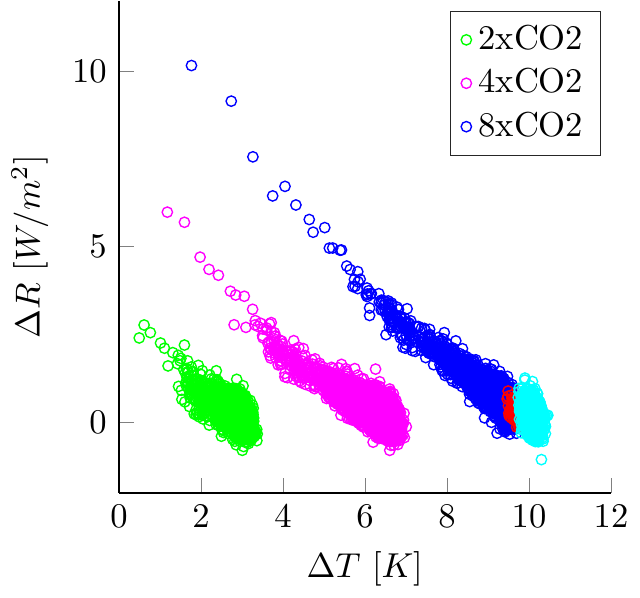}
		\caption{Gregory plot}
	\end{subfigure}
	
	\begin{subfigure}[t]{0.45\textwidth}
		\includegraphics[width=\textwidth]{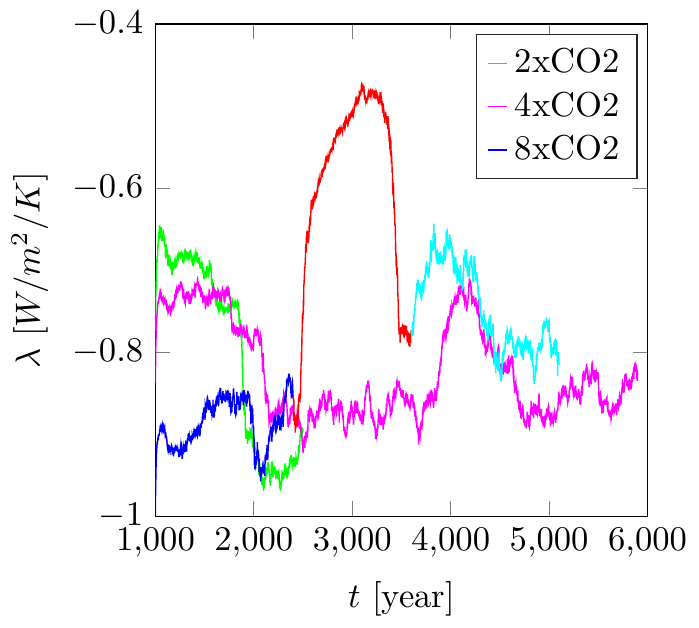}
		\caption{feedback parameter}
	\end{subfigure}
	\begin{subfigure}[t]{0.45\textwidth}
		\includegraphics[width=\textwidth]{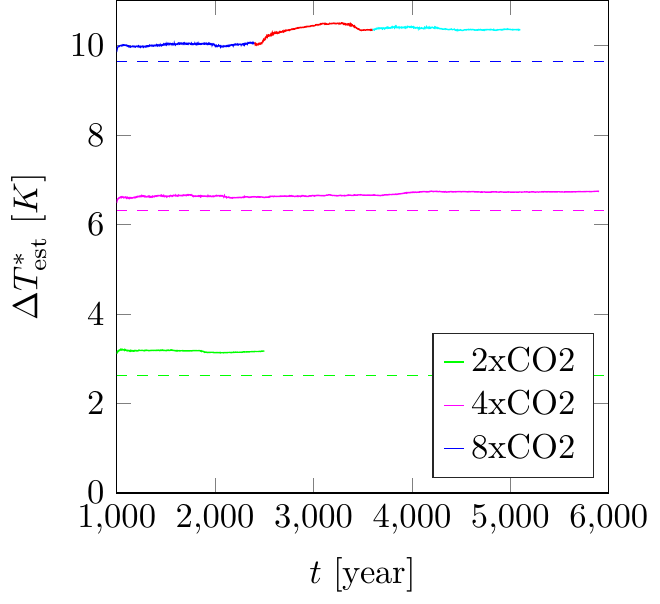}
		\caption{estimated equilibrium warming}
	\end{subfigure}

	\caption{Outcomes of multi-millennial experiments in the GCM `CESM 1.0.4.' for an abrupt2xCO2 (green), an abrupt4xCO2 (magenta) and an abrupt8xCO2 (blue \& red) experiment (Data from longrunmip~\cite{rugenstein2019longrunmip}). For the abrupt8xCO2 experiment, a sudden late increase in temperature can be seen around year $2500$. (a) Time series of global mean surface temperature. (b) Gregory plot. (c) Results for estimated climate feedback parameter $\lambda$ obtained via Gregory fits on time windows of $1,000$ years. (d) Results of estimated equilibrium temperature $\Delta T_\mathrm{est}^*$ via Gregory fits on time windows of $1,000$ years, and estimated warming from a Gregory fit on years 20-150 (dashed lines). For the abrupt8xCO2 experiment the dark respectively light blue indicates years before respectively after the late fast warming. The red data points in (a-b) indicate the time period of the late warming; in (c-d) the red data points indicate that the regression has used some of the data of this late warming period.}
	\label{fig:longrunmip}
\end{figure}

The response of GCMs can also include late rapid changes. An example of such a late warming event is visible around year $2,300$ of the abrupt8xCO2 run in the model CESM 1.0.4 within LongRunMIP~\cite{rugenstein2019longrunmip}. Figure~\ref{fig:longrunmip} shows features of this run, along with associated abrupt2xCO2 and abrupt4xCO2 runs of the same model for comparison. In (a), the time series for the increase in (yearly averaged) global mean near-surface temperature is shown. For the abrupt8xCO2 experiment a late and sudden increase can be seen around $t = 2300$ years (highlighted in red in the figure), which is not present in the other experiments. We have analysed this data using the Gregory method on millennia-long rolling windows (to suppress the natural variability on shorter time scales) in (c-d). We found an increase in the feedback parameter $\lambda$ around the same time, and also an underestimation of the equilibrium warming for $t < 2400$ years. This is similar to our findings in a conceptual energy balance model (Figure~\ref{fig:fastslowGEBM}) albeit less distinct. Hence, we suggest that this late warming event in the abrupt8xCO2 run could be an example of a late tipping event in a GCM. Appendix~\ref{app:CESM8xCO2} illustrates that this tipping behaviour is probably due to a qualitative regional tipping of the AMOC which appears for the 8xCO2 run, but is not present in the 2xCO2 or 4xCO2 runs. However, we note that unlike in Figure~\ref{fig:fastslowGEBM}, the tipping for the 8xCO2 run is of transient nature: the final state is an ``AMOC on'' state in all cases.

\section{Conclusion and discussion}
\label{sec:Discuss}

Although many authors have pointed out deficiencies with estimating and using equilibrium climate sensitivity (ECS), it clearly remains an important metric for understanding the response of climate models to changes in forcing \COO. In particular, although there may be problems with timescales, low frequency variability and lack of linearity, ECS and variants of it are key metrics that find their way (for example, via integrated assessment models of the socioeconomic impact of an emissions pathway such as used in \cite{ipcc-wgII, van2020costs, nordhaus1992dice}) into decision making about climate change and its likely impact on human activities. In this paper, we have illustrated how such linear concepts could break down in many different ways, even after long transient periods in which they seem valid, when nonlinear dynamics start to play a role. Although we have focused in this paper on climate response to idealised abrupt \COO\ forcing scenarios, we also want to stress that in multistable nonlinear systems the precise outcome can also depend on the pathway taken -- that is, not only the amount of emissions but also the moment of emissions can be important. This further complicates and challenges too simplistic linear frameworks. See for instance \cite{ashwin2012tipping,Ritchie.2021} for examples in conceptual settings, as well as \cite{grubb2021modeling} for a discussion on how this can strongly influence integrated assessments.

Climate is a multiscale process that takes place on many fast and slow time scales, so it is unrealistic to assume that all dynamics can be modelled by a univariate linear model. Moreover, we also cannot expect to estimate processes that take place over substantially longer time scales than the simulated duration. As we have illustrated in this paper, this means that even in the case of pure linear response, ECS cannot be accurately estimated unless the simulation times are long enough to resolve the slow timescales such as those common in large scale ocean dynamics or land ice sheets. On top of that, with the examples in Section~\ref{sec:GEBM} we have illustrated how nonlinear effects in a multiscale climate model can lead to additional warming effects -- such as slow and late tipping, with long transients without obvious hints of these late events. These examples demonstrate that even if a fit is very good for a long period of time, there still may be large and abrupt late tipping points.

In section~\ref{sec:EqResponse} and in Figure~\ref{fig:tradeoff5}, we have introduced several trade-offs that need to be made when estimating ECS for a climate model. It would be of great interest to locate the ``Goldilocks Zone'' in which reliable and accurate estimates ECS are possible, in order to give suitable protocols for experiments with GCMs. In particular, it would be good to understand (a) the minimum times and ensemble sizes needed to reliably estimate ECS and (b) the thresholds in perturbation size for general GCMs that lead to tipping behaviour. This will depend not just on the current climate state but also on the processes that are included in the model and the form of the forcing. We suggest there is a need to find criteria that imply that an estimation protocol will work -- and on which time scales. For instance, the Gregory method when applied on data from one decade can typically predict a few decades but is unlikely to be predictive on the scale of centuries; similarly, if only 150 years of data is available, it is unlikely to obtain an accurate estimation on millennial time scales.

The most drastic examples of nonlinear response given in this paper concern tipping phenomena. This begs the question of how relevant this is for future projections with GCMs. After all, in these models the GMST response is typically fairly linear to changes in forcing levels, and the transient response seems linear over quite long timescales. This might suggest that tipping points for GMST are not very relevant. However, the parameter space of such models has not been sufficiently explored to capture past and future tipping~\cite{hopcroft2021paleoclimate} and, consequently, under standard settings (optimised for stable Holocene pre-industrial climates) those GCMs may operate in a too stable manner \cite{Valdes.2011}. Simultaneously, local or regional tipping has been observed more frequently in GCMs \cite{Drijfhout.2015}, and can be observed in past climate records \cite{anna2020quantification}. Tipping effects at regional levels may give only a small signal in the global average (although e.g. the AMOC restoration in the abrupt8xCO2 experiment in Figure~\ref{fig:longrunmip} is visible in GMST). Indeed, one might conjecture a global redistribution that almost averages out in data of GMST -- similar to what is described in~\cite{rietkerk2021evasion, bastiaansen2022fragmented}. However, such regional tipping is much more problematic than a global mean signal might indicate, as local impacts can be very dramatic. Moreover, when several regional tipping elements are involved, {\em cascading} effects may occur \cite{Dekker.2018, Wunderling.2021} opening the possibility of an eventual global response for example through triggering of additional carbon cycle feedbacks. It remains an important issue for future climate projections to determine which tipping points may be crossed on time scales of centuries to millennia. It also highlights the importance of going beyond classifying climate response only via GMST, to look at spatial responses and other observables.

\enlargethispage{20pt}

\section*{Data statement}
Simulation data from models in LongRunMIP \url{data.iac.ethz.ch/longrunmip/}, including the here used model CESM 1.0.4, requests for access can be made to the coordinators of longrunMIP. More information and details of the simulations can be found on \url{longrunmip.org} and in \cite{rugenstein2019longrunmip}.\\
The numerical code to simulate and subsequently analyse the conceptual energy balance model introduced in equations \eqref{eq:0DGEBM}, \eqref{eq:fsalpha}, \eqref{eq:NV} is available from \url{https://github.com/peterashwin/late-tipping-2022}

\section*{Author contributions}
All authors designed the study. RB and PA undertook the computer simulations and analysis. All authors edited the final text.

\section*{Acknowledgements}
We thank Richard Wood for discussions related to this work. This project is TiPES contribution \# 168: This project has received funding from the European Union’s Horizon 2020 research and innovation programme under Grant Agreement 820970.

\bibliographystyle{unsrt}
\bibliography{refs}

\begin{thebibliography}{10}

\bibitem{Arrhenius1896}
Svante Arrhenius.
\newblock On the influence of carbonic acid in the air upon the temperature of
  the ground.
\newblock {\em The London, Edinburgh, and Dublin Philosophical Magazine and
  Journal of Science}, 41(251):237--276, 1896.

\bibitem{aboutArrhenius}
Andrei~G. Lapenis.
\newblock Arrhenius and the intergovernmental panel on climate change.
\newblock {\em Eos, Transactions American Geophysical Union}, 79(23):271--271,
  1998.

\bibitem{Charney.1979}
J~G Charney.
\newblock {\em {Carbon Dioxide and Climate: A Scientific Assessment}}.
\newblock National Academy of Science. National Academy of Science, 1979.

\bibitem{Sherwood.2020}
S~Sherwood, Mark~J Webb, J~D Annan, Kyle~C Armour, Piers~M Forster, J~C
  Hargreaves, Gabi Hegerl, S~A Klein, K~D Marvel, E~J Rohling, M~Watanabe,
  Timothy Andrews, P~Braconnot, C~S Bretherton, Gavin~L Foster, Z~Hausfather,
  Anna~S von~der Heydt, Reto Knutti, T~Mauritsen, J~R Norris, C~Proistosescu,
  M~Rugenstein, G~A Schmidt, K~B Tokarska, and M~D Zelinka.
\newblock {An assessment of Earth's climate sensitivity using multiple lines of
  evidence}.
\newblock {\em Reviews of Geophysics}, page e2019RG000678, 2020.

\bibitem{senior2000time}
Catherine~A Senior and John~FB Mitchell.
\newblock The time-dependence of climate sensitivity.
\newblock {\em Geophysical Research Letters}, 27(17):2685--2688, 2000.

\bibitem{knutti2015feedbacks}
Reto Knutti and Maria~AA Rugenstein.
\newblock Feedbacks, climate sensitivity and the limits of linear models.
\newblock {\em Philosophical Transactions of the Royal Society A: Mathematical,
  Physical and Engineering Sciences}, 373(2054):20150146, 2015.

\bibitem{eyring2016overview}
Veronika Eyring, Sandrine Bony, Gerald~A Meehl, Catherine~A Senior, Bjorn
  Stevens, Ronald~J Stouffer, and Karl~E Taylor.
\newblock Overview of the coupled model intercomparison project phase 6 (cmip6)
  experimental design and organization.
\newblock {\em Geoscientific Model Development}, 9(5):1937--1958, 2016.

\bibitem{rugenstein2019longrunmip}
Maria Rugenstein, Jonah Bloch-Johnson, Ayako Abe-Ouchi, Timothy Andrews, Urs
  Beyerle, Long Cao, Tarun Chadha, Gokhan Danabasoglu, Jean-Louis Dufresne, Lei
  Duan, et~al.
\newblock Longrunmip: motivation and design for a large collection of
  millennial-length aogcm simulations.
\newblock {\em Bulletin of the American Meteorological Society},
  100(12):2551--2570, 2019.

\bibitem{ashwin2020extreme}
Peter Ashwin and Anna~S. von~der Heydt.
\newblock Extreme sensitivity and climate tipping points.
\newblock {\em Journal of Statistical Physics}, 179(5):1531--1552, 2020.

\bibitem{von2016state}
Anna~S von~der Heydt and Peter Ashwin.
\newblock State dependence of climate sensitivity: attractor constraints and
  palaeoclimate regimes.
\newblock {\em Dynamics and Statistics of the Climate System}, 1(1):dzx001,
  2016.

\bibitem{eckmann1985ergodic}
J-P Eckmann and David Ruelle.
\newblock Ergodic theory of chaos and strange attractors.
\newblock {\em The theory of chaotic attractors}, pages 273--312, 1985.

\bibitem{young2017generalizations}
Lai-Sang Young.
\newblock Generalizations of srb measures to nonautonomous, random, and
  infinite dimensional systems.
\newblock {\em Journal of Statistical Physics}, 166(3-4):494--515, 2017.

\bibitem{young2002srb}
Lai-Sang Young.
\newblock What are srb measures, and which dynamical systems have them?
\newblock {\em Journal of Statistical Physics}, 108(5):733--754, 2002.

\bibitem{Rugenstein2021ThreeFlavors}
Maria A.~A. Rugenstein and Kyle~C. Armour.
\newblock Three flavors of radiative feedbacks and their implications for
  estimating equilibrium climate sensitivity.
\newblock {\em Geophysical Research Letters}, 48(15):e2021GL092983, 2021.
\newblock e2021GL092983 2021GL092983.

\bibitem{Caballero.2013}
Rodrigo Caballero and M~Huber.
\newblock {State-dependent climate sensitivity in past warm climates and its
  implications for future climate projections}.
\newblock {\em Proceedings of the National Academy of Science}, 110(35):14162
  -- 14167, 2013.

\bibitem{rugenstein2020equilibrium}
Maria Rugenstein, Jonah Bloch-Johnson, Jonathan Gregory, Timothy Andrews,
  Thorsten Mauritsen, Chao Li, Thomas~L. Frölicher, David Paynter, Gokhan
  Danabasoglu, Shuting Yang, Jean-Louis Dufresne, Long Cao, Gavin~A. Schmidt,
  Ayako Abe-Ouchi, Olivier Geoffroy, and Reto Knutti.
\newblock Equilibrium climate sensitivity estimated by equilibrating climate
  models.
\newblock {\em Geophysical Research Letters}, 47(4):e2019GL083898, 2020.
\newblock e2019GL083898 10.1029/2019GL083898.

\bibitem{gregory2004new}
JM~Gregory, WJ~Ingram, MA~Palmer, GS~Jones, PA~Stott, RB~Thorpe, JA~Lowe,
  TC~Johns, and KD~Williams.
\newblock A new method for diagnosing radiative forcing and climate
  sensitivity.
\newblock {\em Geophysical research letters}, 31(3), 2004.

\bibitem{Bastiaansen:2021fm}
Robbin Bastiaansen, Henk~A. Dijkstra, and Anna~S. von~der Heydt.
\newblock {Multivariate Estimations of Equilibrium Climate Sensitivity From
  Short Transient Warming Simulations}.
\newblock {\em Geophysical Research Letters}, 48(1):e2020GL091090, 2021.

\bibitem{andrews2015dependence}
Timothy Andrews, Jonathan~M Gregory, and Mark~J Webb.
\newblock The dependence of radiative forcing and feedback on evolving patterns
  of surface temperature change in climate models.
\newblock {\em Journal of Climate}, 28(4):1630--1648, 2015.

\bibitem{knutti2017beyond}
Reto Knutti, Maria~AA Rugenstein, and Gabriele~C Hegerl.
\newblock Beyond equilibrium climate sensitivity.
\newblock {\em Nature Geoscience}, 10(10):727--736, 2017.

\bibitem{Cummins.2020}
Donald~P. Cummins, David~B. Stephenson, and Peter~A. Stott.
\newblock Optimal estimation of stochastic energy balance model parameters.
\newblock {\em Journal of Climate}, 33(18):7909 -- 7926, 2020.

\bibitem{dai2020improved}
Aiguo Dai, Danqing Huang, Brian~EJ Rose, Jian Zhu, and Xiangjun Tian.
\newblock Improved methods for estimating equilibrium climate sensitivity from
  transient warming simulations.
\newblock {\em Climate Dynamics}, 54(11):4515--4543, 2020.

\bibitem{geoffroy2013transient}
Olivier Geoffroy, D~Saint-Martin, G~Bellon, A~Voldoire, DJL Olivi{\'e}, and
  S~Tyt{\'e}ca.
\newblock Transient climate response in a two-layer energy-balance model. part
  ii: Representation of the efficacy of deep-ocean heat uptake and validation
  for cmip5 aogcms.
\newblock {\em Journal of Climate}, 26(6):1859--1876, 2013.

\bibitem{Lucarini.2018}
Valerio Lucarini.
\newblock {Revising and Extending the Linear Response Theory for Statistical
  Mechanical Systems: Evaluating Observables as Predictors and Predictands}.
\newblock {\em Journal of Statistical Physics}, 173(6):1698--1721, 2018.

\bibitem{ruelle2009review}
David Ruelle.
\newblock A review of linear response theory for general differentiable
  dynamical systems.
\newblock {\em Nonlinearity}, 22(4):855, 2009.

\bibitem{Ragone.2016}
F~Ragone, Valerio Lucarini, and F~Lunkeit~Climate Dynamics.
\newblock {A new framework for climate sensitivity and prediction: a modelling
  perspective}.
\newblock {\em Climate Dynamics}, 46(5-6):1459 -- 1471, 2016.

\bibitem{lucarini2011statistical}
Valerio Lucarini and Stefania Sarno.
\newblock A statistical mechanical approach for the computation of the climatic
  response to general forcings.
\newblock {\em Nonlinear Processes in Geophysics}, 18(1):7--28, 2011.

\bibitem{proistosescu2017slow}
Cristian Proistosescu and Peter~J Huybers.
\newblock Slow climate mode reconciles historical and model-based estimates of
  climate sensitivity.
\newblock {\em Science advances}, 3(7):e1602821, 2017.

\bibitem{hasselmann1993cold}
Klaus Hasselmann, Robert Sausen, Ernst Maier-Reimer, and Reinhard Voss.
\newblock On the cold start problem in transient simulations with coupled
  atmosphere-ocean models.
\newblock {\em Climate Dynamics}, 9(2):53--61, 1993.

\bibitem{lembo2020beyond}
Valerio Lembo, Valerio Lucarini, and Francesco Ragone.
\newblock Beyond forcing scenarios: predicting climate change through response
  operators in a coupled general circulation model.
\newblock {\em Scientific Reports}, 10(1):1--13, 2020.

\bibitem{aengenheyster2018point}
Matthias Aengenheyster, Qing~Yi Feng, Frederick Van Der~Ploeg, and Henk~A
  Dijkstra.
\newblock The point of no return for climate action: effects of climate
  uncertainty and risk tolerance.
\newblock {\em Earth System Dynamics}, 9(3):1085--1095, 2018.

\bibitem{bastiaansen2021projections}
Robbin Bastiaansen, Henk~A Dijkstra, and Anna S von~der Heydt.
\newblock Projections of the transient state-dependency of climate feedbacks.
\newblock {\em Geophysical Research Letters}, 48(20):e2021GL094670, 2021.

\bibitem{torres2021identification}
Guilherme~L Torres~Mendon{\c{c}}a, Julia Pongratz, and Christian~H Reick.
\newblock Identification of linear response functions from arbitrary
  perturbation experiments in the presence of noise--part 1: Method development
  and toy model demonstration.
\newblock {\em Nonlinear Processes in Geophysics}, 28(4):501--532, 2021.

\bibitem{maier1987transport}
Ernst Maier-Reimer and Klaus Hasselmann.
\newblock Transport and storage of co2 in the ocean——an inorganic
  ocean-circulation carbon cycle model.
\newblock {\em Climate dynamics}, 2(2):63--90, 1987.

\bibitem{anna2020quantification}
Anna~S. von~der Heydt, Peter Ashwin, Charles~D Camp, Michel Crucifix, Henk~A
  Dijkstra, Peter Ditlevsen, and Timothy~M Lenton.
\newblock Quantification and interpretation of the climate variability record.
\newblock {\em Global and Planetary Change}, page 103399, 2020.

\bibitem{mitchell1976overview}
J~Murray Mitchell.
\newblock An overview of climatic variability and its causal mechanisms.
\newblock {\em Quaternary Research}, 6(4):481--493, 1976.

\bibitem{Baatsen:2020vz}
Michiel L~J Baatsen, Anna~S von~der Heydt, M~Huber, Michael~A Kliphuis, Peter~K
  Bijl, Appy Sluijs, and Henk~A. Dijkstra.
\newblock {The middle-to-late Eocene greenhouse climate, modelled using the
  CESM 1.0.5}.
\newblock {\em Climate of the Past}, 16(6):2573--2597, 2020.

\bibitem{budyko}
M.~I. Budyko.
\newblock The effect of solar radiation variations on the climate of the earth.
\newblock {\em Tellus}, 21(5):611--619, 1969.

\bibitem{sellers}
William~D. Sellers.
\newblock A global climatic model based on the energy balance of the
  earth-atmosphere system.
\newblock {\em Journal of Applied Meteorology and Climatology}, 8(3):392 --
  400, 1969.

\bibitem{ghil1976climate}
Michael Ghil.
\newblock Climate stability for a sellers-type model.
\newblock {\em Journal of Atmospheric Sciences}, 33(1):3--20, 1976.

\bibitem{Myhre2013}
G~Myhre, D~Shindell, F-M Bréon, W~Collins, J~Fuglestvedt, J~Huang, D~Koch,
  Lamargque J.-F, D~Lee, B~Mendoza, T~Nakajima, A~Robock, G~Stephens,
  T~Takemura, and H~Zhang.
\newblock Antopogenic and natural radiative forcing.
\newblock In T.~F. Stocker, D.~Qin, G.-K. Plattner, M.~Tignor, S.K. Allen,
  J.~Boschung, A~Nauels, Y.~Xia, V.~Bex, and P.M. Midgley, editors, {\em
  Climate Change 2013: The Physical Science Basis. Contribution of Working
  Group I to the Fifth Assessment Report on the Intergovernmental Panel on
  Climate Change}, chapter~8, pages 659--740. Cambridge University Press,
  Cambridge, United Kingdom and New York, NY, USA, 2013.

\bibitem{lorenz1963deterministic}
Edward~N Lorenz.
\newblock Deterministic nonperiodic flow.
\newblock {\em Journal of atmospheric sciences}, 20(2):130--141, 1963.

\bibitem{montaldi2021}
James Montaldi.
\newblock {\em Singularities, Bifurcations and Catastrophes}.
\newblock Cambridge University Press, 2021.

\bibitem{Heydt.2016}
Anna~S von~der Heydt, Henk~A. Dijkstra, Roderik S W Van~De Wal, Rodrigo
  Caballero, Michel Crucifix, Gavin~L Foster, M~Huber, Peter Koehler, E~J
  Rohling, Paul~J Valdes, Peter Ashwin, Sebastian Bathiany, T~Berends, L~van
  Bree, Peter~D Ditlevsen, Michael Ghil, Alan~M Haywood, Joel Katzav, Gerrit
  Lohmann, J~Lohmann, Valerio Lucarini, A~Marzocchi, H~P alike, I~Ruvalcaba
  Baroni, D~Simon, Appy Sluijs, L~B Stap, A~Tantet, J~P Viebahn, and Martin
  Ziegler.
\newblock {Lessons on Climate Sensitivity From Past Climate Changes}.
\newblock {\em Current Climate Change Reports}, 2(4):148 -- 158, 2016.

\bibitem{alkhayuon2018rate}
Hassan~M Alkhayuon and Peter Ashwin.
\newblock Rate-induced tipping from periodic attractors: Partial tipping and
  connecting orbits.
\newblock {\em Chaos: An Interdisciplinary Journal of Nonlinear Science},
  28(3):033608, 2018.

\bibitem{ashwin2021physical}
Peter Ashwin and Julian Newman.
\newblock Physical invariant measures and tipping probabilities for chaotic
  attractors of asymptotically autonomous systems.
\newblock {\em The European Physical Journal Special Topics},
  230(16):3235--3248, 2021.

\bibitem{stouffer2006investigating}
Ronald~J Stouffer, J~Yin, JM~Gregory, KW~Dixon, MJ~Spelman, W~Hurlin,
  AJ~Weaver, M~Eby, GM~Flato, H~Hasumi, et~al.
\newblock Investigating the causes of the response of the thermohaline
  circulation to past and future climate changes.
\newblock {\em Journal of climate}, 19(8):1365--1387, 2006.

\bibitem{jackson2015global}
LC~Jackson, R~Kahana, T~Graham, MA~Ringer, T~Woollings, JV~Mecking, and
  RA~Wood.
\newblock Global and european climate impacts of a slowdown of the amoc in a
  high resolution gcm.
\newblock {\em Climate dynamics}, 45(11):3299--3316, 2015.

\bibitem{Ritchie.2021}
Paul D~L Ritchie, Joseph~J Clarke, Peter~M Cox, and Chris Huntingford.
\newblock {Overshooting tipping point thresholds in a changing climate}.
\newblock {\em Nature}, 592(7855):517 -- 523, 2021.

\bibitem{deser2020insights}
Clara Deser, Flavio Lehner, Keith~B Rodgers, Toby Ault, Thomas~L Delworth,
  Pedro~N DiNezio, Arlene Fiore, Claude Frankignoul, John~C Fyfe, Daniel~E
  Horton, et~al.
\newblock Insights from earth system model initial-condition large ensembles
  and future prospects.
\newblock {\em Nature Climate Change}, 10(4):277--286, 2020.

\bibitem{aragon2016mip}
MA~Aragon-Calvo.
\newblock The mip ensemble simulation: local ensemble statistics in the cosmic
  web.
\newblock {\em Monthly Notices of the Royal Astronomical Society},
  455(1):438--448, 2016.

\bibitem{manabe1993century}
Syukuro Manabe and Ronald~J Stouffer.
\newblock Century-scale effects of increased atmospheric c02 on the
  ocean--atmosphere system.
\newblock {\em Nature}, 364(6434):215--218, 1993.

\bibitem{manabe1994multiple}
Syukuro Manabe and Ronald~J Stouffer.
\newblock Multiple-century response of a coupled ocean-atmosphere model to an
  increase of atmospheric carbon dioxide.
\newblock {\em Journal of climate}, 7(1):5--23, 1994.

\bibitem{ipcc-wgII}
H.-O. Pörtner, D.C. Roberts, M.~Tignor, E.S. Poloczanska, K.~Mintenbeck,
  A.~Alegría, M.~Craig, S.~Langsdorf, S.~Löschke, V.~Möller, A.~Okem, and
  B.~Rama.
\newblock {\em IPCC, 2022: Climate Change 2022: Impacts, Adaptation, and
  Vulnerability. Contribution of Working Group II to the Sixth Assessment
  Report of the Intergovernmental Panel on Climate Change}.
\newblock Cambridge University Press, In Press.

\bibitem{van2020costs}
Detlef~P van Vuuren, Kaj-Ivar van~der Wijst, Stijn Marsman, Maarten van~den
  Berg, Andries~F Hof, and Chris~D Jones.
\newblock The costs of achieving climate targets and the sources of
  uncertainty.
\newblock {\em Nature Climate Change}, 10(4):329--334, 2020.

\bibitem{nordhaus1992dice}
William Nordhaus.
\newblock The 'dice' model: Background and structure of a dynamic integrated
  climate-economy model of the economics of global warming.
\newblock Cowles Foundation Discussion Papers 1009, Cowles Foundation for
  Research in Economics, Yale University, 1992.

\bibitem{ashwin2012tipping}
Peter Ashwin, Sebastian Wieczorek, Renato Vitolo, and Peter Cox.
\newblock Tipping points in open systems: bifurcation, noise-induced and
  rate-dependent examples in the climate system.
\newblock {\em Philosophical Transactions of the Royal Society A: Mathematical,
  Physical and Engineering Sciences}, 370(1962):1166--1184, 2012.

\bibitem{grubb2021modeling}
Michael Grubb, Claudia Wieners, and Pu~Yang.
\newblock Modeling myths: On dice and dynamic realism in integrated assessment
  models of climate change mitigation.
\newblock {\em Wiley Interdisciplinary Reviews: Climate Change}, 12(3):e698,
  2021.

\bibitem{hopcroft2021paleoclimate}
Peter~O Hopcroft and Paul~J Valdes.
\newblock Paleoclimate-conditioning reveals a north africa land--atmosphere
  tipping point.
\newblock {\em Proceedings of the National Academy of Sciences},
  118(45):e2108783118, 2021.

\bibitem{Valdes.2011}
Paul Valdes.
\newblock {Built for stability}.
\newblock {\em Nature Geoscience}, 4(7):414 -- 416, 2011.

\bibitem{Drijfhout.2015}
Sybren Drijfhout, Sebastian Bathiany, Claudie Beaulieu, Victor Brovkin, Martin
  Claussen, Chris Huntingford, Marten Scheffer, Giovanni Sgubin, and Didier
  Swingedouw.
\newblock {Catalogue of abrupt shifts in Intergovernmental Panel on Climate
  Change climate models.}
\newblock {\em Proceedings of the National Academy of Sciences of the United
  States of America}, 112(43):E5777 -- 86, 2015.

\bibitem{rietkerk2021evasion}
Max Rietkerk, Robbin Bastiaansen, Swarnendu Banerjee, Johan van~de Koppel, Mara
  Baudena, and Arjen Doelman.
\newblock Evasion of tipping in complex systems through spatial pattern
  formation.
\newblock {\em Science}, 374(6564):eabj0359, 2021.

\bibitem{bastiaansen2022fragmented}
Robbin Bastiaansen, Henk~A Dijkstra, and Anna~S von~der Heydt.
\newblock Fragmented tipping in a spatially heterogeneous world.
\newblock {\em Environmental Research Letters}, 17(4):045006, 2022.

\bibitem{Dekker.2018}
Mark~M Dekker, Anna~S von~der Heydt, and Henk~A. Dijkstra.
\newblock {Cascading transitions in the climate system}.
\newblock {\em Earth System Dynamics}, 9:1243 -- 1260, 2018.

\bibitem{Wunderling.2021}
Nico Wunderling, Jonathan~F. Donges, Jürgen Kurths, and Ricarda Winkelmann.
\newblock {Interacting tipping elements increase risk of climate domino effects
  under global warming}.
\newblock {\em Earth System Dynamics}, 12(2):601--619, 2021.

\bibitem{stewart1980}
Ian~N. Stewart.
\newblock Catastrophe theory and equations of state: conditions for a butterfly
  singularity.
\newblock {\em Mathematical Proceedings of the Cambridge Philosophical
  Society}, 88(3):429–449, 1980.

\bibitem{dhooge2008new}
Annick Dhooge, Willy Govaerts, Yu~A Kuznetsov, Hil Ga{\'e}tan~Ellart Meijer,
  and Bart Sautois.
\newblock New features of the software matcont for bifurcation analysis of
  dynamical systems.
\newblock {\em Mathematical and Computer Modelling of Dynamical Systems},
  14(2):147--175, 2008.

\end{thebibliography}

\appendix

\clearpage
\newpage

\section{Bifurcation structure of the GEBM}
\label{app:butterfly}

In the absence of chaotic forcing ($\nu_{NV}\equiv 0$), the energy balance model~\eqref{eq:0DGEBM} has equilibria that are independent of the value of $\tau_\alpha\geq 0$; hence we study here the equilibria for the case of no dynamic albedo. That is, we consider
\begin{equation}
   C \frac{dT}{dt} = Q_0 (1 - \alpha_0(T)) - \varepsilon_0(T) \sigma T^4 + \mu.
   \label{eq:appGEBM}
\end{equation}
with $\alpha_0$ and $\varepsilon_0$ as in (\ref{eq:alpha0}),(\ref{eq:varepsilon}).
To study the bifurcation structure of this equation, we apply the scalings
$y:= K_\alpha T$ and $s:= \sigma (\varepsilon_1+\varepsilon_2)/(2 K_a^3C) t$, and we introduce the following (composed) parameters
\begin{equation}
{ \everymath={\displaystyle}
\begin{array}{rlrl}
	\nu & := \frac{K_\alpha^4}{\sigma} \frac{Q_0 \left(2 - \alpha_1 - \alpha_2\right) + 2 \mu}{\varepsilon_1+\varepsilon_2}~~&
	a & := \frac{Q_0 K_\alpha^4}{\sigma} \frac{\alpha_1-\alpha_2}{\varepsilon_1+\varepsilon_2} \\
	c & := \frac{\varepsilon_1-\varepsilon_2}{\varepsilon_1+\varepsilon_2} &
	d & := \frac{K_\varepsilon}{K_\alpha}; \\
	y_\alpha & := K_\alpha T_\alpha; &
	y_\varepsilon & := K_\alpha T_\varepsilon.
\end{array}
}
\end{equation}
Then \eqref{eq:appGEBM} becomes
\begin{equation}
	\frac{dy}{ds} = f(y) := \nu + a \tanh\left[y-y_\alpha\right] - y^4 \left( 1 - c \tanh\left[ d\left(y-y_\varepsilon\right)\right]\right)
\end{equation}
This system has an equilibrium at $y = y_r$ when $f(y_r) = 0$ for a given set of parameter values. However, bifurcations can occur as parameters change. At such bifurcation points, not only $f(y_r) = 0$, but also derivatives of $f$ with respect to $y$ vanish. How many derivatives vanish denotes the co-dimension and degeneracy of the bifurcation. If only the first derivative vanishes, it is a saddle-node bifurcation in which two equilibria collide and disappear. If the first two derivatives vanish, it is a cusp bifurcation, in which three equilibria meet -- or, in other words, two saddle-node bifurcations. If the first three derivatives vanish, it is called a swallowtail point, at which four equilibria meet (or two cusp bifurcations). If the first four derivatives vanish, it is called a butterfly catastrophe or butterfly singularity~\cite{stewart1980,montaldi2021}, where five equilibria or two swallowtail points meet (or three cusp bifurcations, or four saddle-node bifurcations).

To study the bifurcation structure, it is therefore useful to look at the Taylor expansion of $f$ around a reference point $y_r$. For this we set $y = y_r + z$ and define $z_\alpha := y_\alpha - y_r$, $z_\varepsilon := y_\varepsilon - y_r$. Then we obtain
\begin{equation}
	f(y_r + z) = \nu + a \tanh\left[ z - z_\alpha \right] - (y + y_r)^4 \left( 1 - c \tanh\left[d(z-z_\varepsilon)\right]\right).
	\label{eq:f}
\end{equation}
Using computer algebra software such as Mathematica the expansion around $z = 0$ of this equation can be computed as
\begin{equation}
	f(y_r + z) = f_0 + f_1 z + f_2 z^2 + f_3 z^3 + f_4 z^4 + f_5 z^5 + \mathcal{O}(z^6)
\end{equation}
with expressions for $f_i$ given in Supplementary Material~\ref{sec:fi}.

\subsection{Cusp bifurcations when $c = 0$ or $a = 0$}

We first inspect some limit cases, starting with the limit case in which $c = 0$. That is, $\varepsilon_1 = \varepsilon_2$, indicating the emissivity does not change with temperature. In this setting, it can be shown that $f_0 = f_1 = f_2 = 0$, whenever the following conditions hold simultaneously:
$$\begin{aligned}
	\nu & = \frac{27}{16} \frac{8 \tanh^2(z_\alpha)-3\sech^2(z_\alpha)}{\sech^2(z_\alpha)\tanh^4(z_\alpha)} \\
	a & = \frac{27}{2} \frac{1}{\sech^2(z_\alpha)\tanh^3(z_\alpha)}\\
	y_r & = \frac{3}{2} \frac{1}{\tanh(z_\alpha)}
\end{aligned}$$
Since it is required that $a \geq 0$ and $y_r \geq 0$, from these expressions it can be seen that cusps bifurcations can only occur if $z_\alpha > 0$. Further, higher order degeneracies cannot occur (no choice for $z_\alpha$ leads to $f_0=f_1=f_2=f_3 = 0$ in the case $c = 0$).

\begin{figure}

\centering

\begin{subfigure}[t]{0.21\textwidth}
	\centering
	\includegraphics[width=\textwidth]{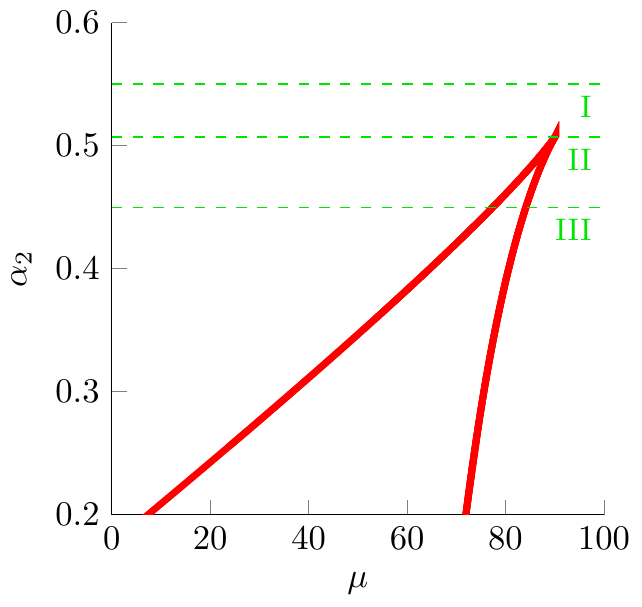}
	\caption{Lines of saddle-node bifurcations}
\end{subfigure}
\begin{subfigure}[t]{0.21\textwidth}
	\centering
	\includegraphics[width=\textwidth]{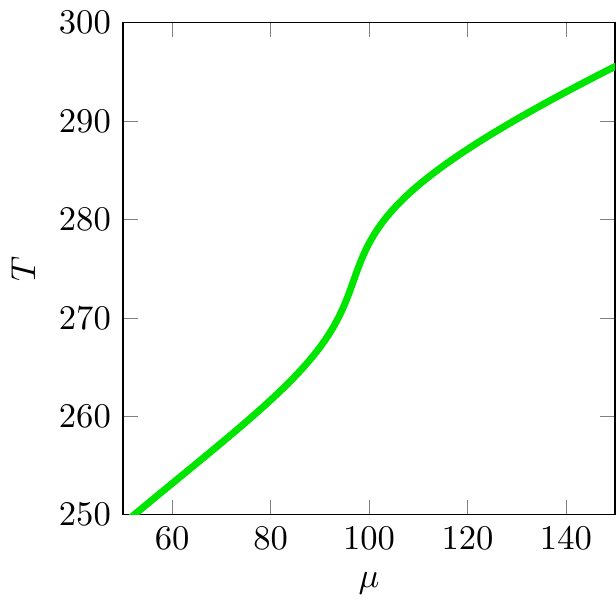}
	\caption{$\alpha_2 = 0.55$}
\end{subfigure}
\begin{subfigure}[t]{0.21\textwidth}
	\centering
	\includegraphics[width=\textwidth]{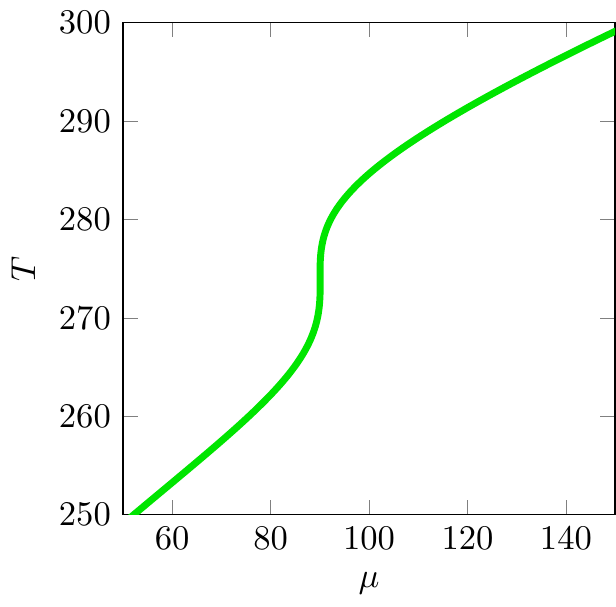}
	\caption{$\alpha_2 = 0.5071$ (cusp)}
\end{subfigure}
\begin{subfigure}[t]{0.21\textwidth}
	\centering
	\includegraphics[width=\textwidth]{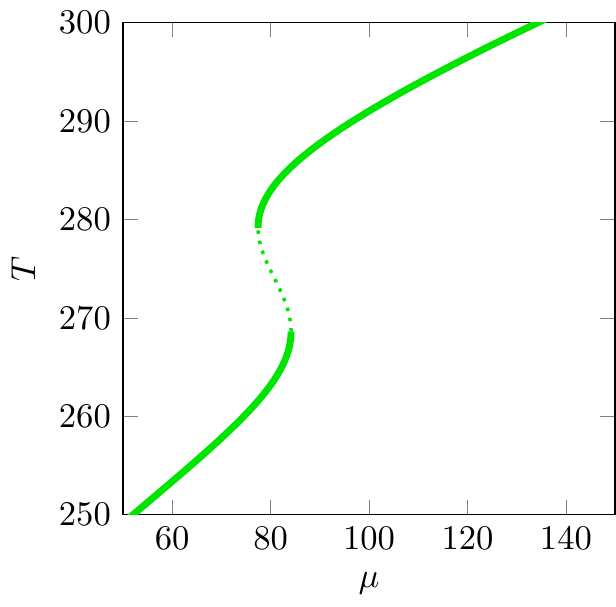}
	\caption{$\alpha_2 = 0.45$}
\end{subfigure}

\caption{Numerical continuation of the saddle-node lines have been performed using Matcont~\cite{dhooge2008new}. (a) The cusp bifurcation in parameter space, with red lines indicating the loci of fold point. Green dashed lines indicate the location in parameter space of the bifurcation diagrams shown in (b-c), where solid green lines indicate stable and dotted green lines indicate unstable branches of equilibria.}
\label{fig:BD_albedo}
\end{figure}

These results can be brought back to the original scaling by a series of substitutions and manipulations. For instance, fixing $Q_0 = 341.3$, $\sigma = 5.67 \cdot 10^{-8}$, $K_a = 0.1$, $T_a = 274.5$, $\varepsilon_1 = \varepsilon_2 = 0.7$, $\alpha_1 = 0.7$ it can be shown that a cusp bifurcation occurs when $\alpha_2 \approx 0.5071$ for $\mu \approx 91.663$ at $T \approx 273.9529$. For lower values of $\alpha_2$, the bifurcation diagram $(\mu,T)$ has two saddle-node bifurcations; for higher values it has no saddle-node bifurcations. See figure~\ref{fig:BD_albedo} for numerical continuation of the saddle-node lines in $(\mu,\alpha_2)$-parameter space, and bifurcation diagrams $(\mu,T)$ for various choices of $\alpha_2$ below, above and at the critical value for which the cusp bifurcation occurs.

Another limit case arises when $a = 0$. In this case, $\alpha_1 = \alpha_2$, indicating that the albedo does not change with temperature. In this setting, it can be shown that $f_0 = f_1 = f_2 = 0$ occurs whenever the following conditions hold:
$$\begin{aligned}
	\nu & = \frac{3125}{16} \frac{1 - \tanh^2(d z_\varepsilon)}{d^4 \tanh^4(d z_\varepsilon) \left(5 + 3 \tanh^2(d z_\varepsilon)\right)} \\
	c & = - \frac{8 \tanh(d z_\varepsilon)}{5 + 3 \tanh^2(d z_\varepsilon)} \\
	y_r & = - \frac{5}{2} \frac{1}{d \tanh(d z_\varepsilon)}
\end{aligned}$$
In this case, to ensure $c \geq 0$ and $y_r > 0$, it is necessary to take $z_\varepsilon < 0$. Observe that higher order degeneracies cannot occur.

\begin{figure}

\centering

\begin{subfigure}[t]{0.21\textwidth}
	\centering
	\includegraphics[width=\textwidth]{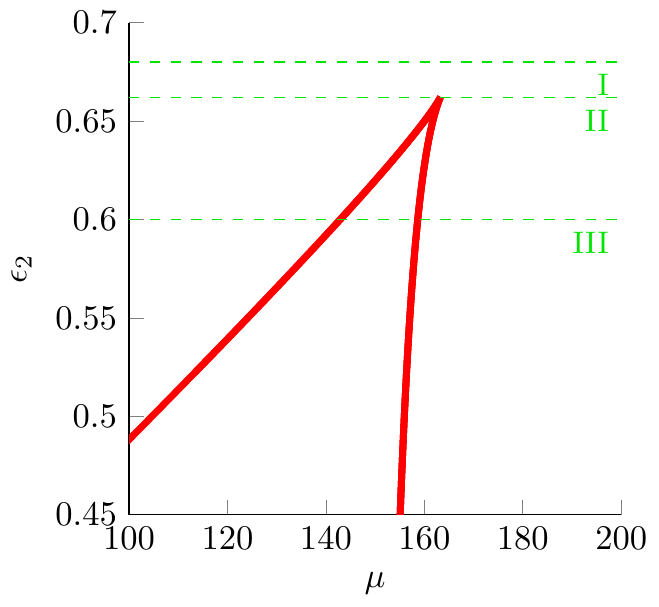}
	\caption{Lines of saddle-node bifurcations}
\end{subfigure}
\begin{subfigure}[t]{0.21\textwidth}
	\centering
	\includegraphics[width=\textwidth]{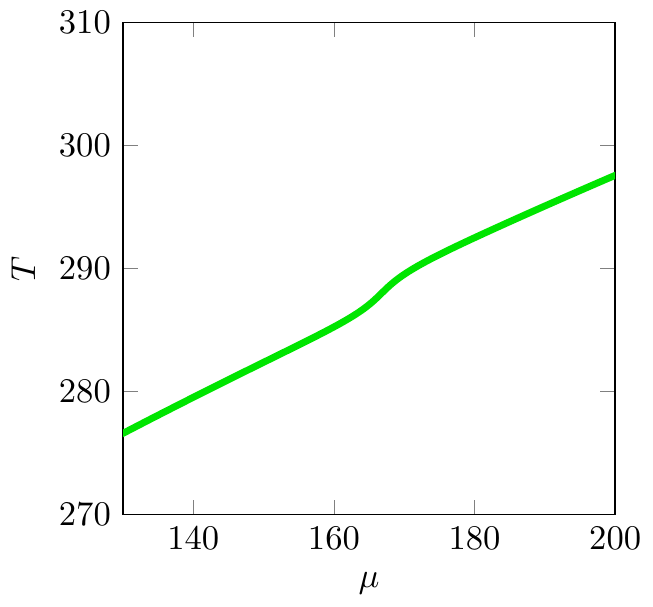}
	\caption{$\varepsilon_2 = 0.68$}
\end{subfigure}
\begin{subfigure}[t]{0.21\textwidth}
	\centering
	\includegraphics[width=\textwidth]{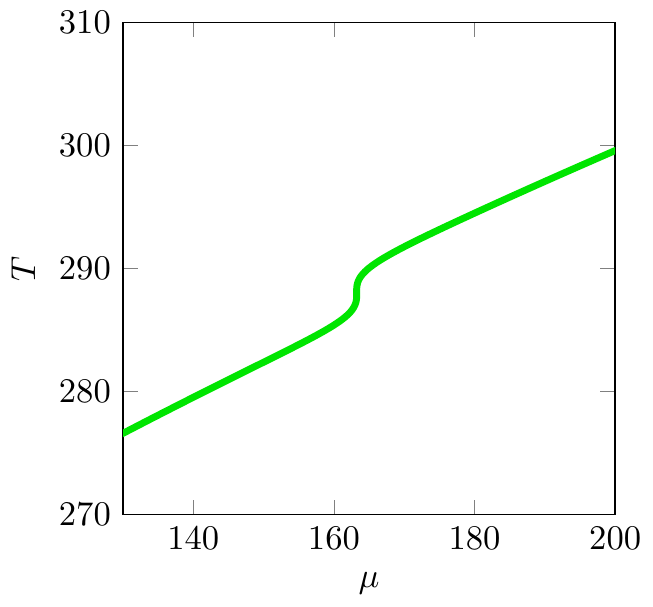}
	\caption{$\varepsilon_2 = 0.6619$}
\end{subfigure}
\begin{subfigure}[t]{0.21\textwidth}
	\centering
	\includegraphics[width=\textwidth]{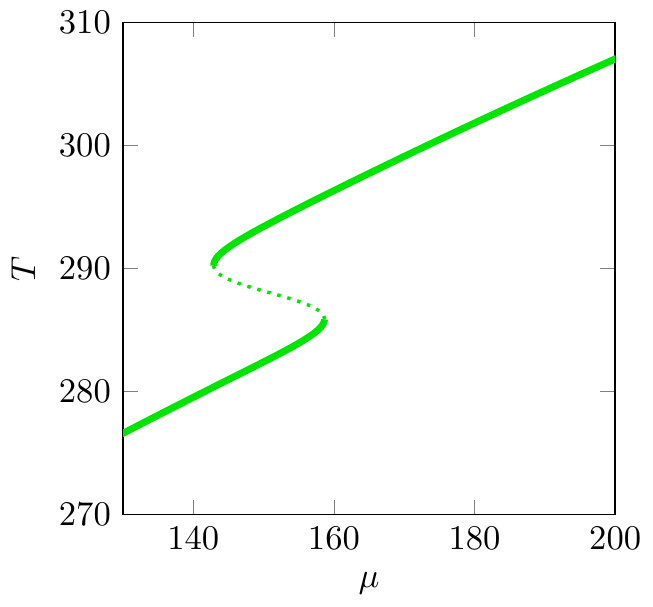}
	\caption{$\varepsilon_2 = 0.6$}
\end{subfigure}

\caption{Numerical continuation of the saddle-node lines have been performed using Matcont~\cite{dhooge2008new}. (a) The cusp bifurcation in parameter space, with red lines indicating the loci of fold points. Green dashed lines indicate the location in parameter space of the bifurcation diagrams shown in (b-c), where solid green lines indicate stable and dotted green lines indicate unstable branches of equilibria.}
\label{fig:BD_emmisivity}
\end{figure}

Again, these results can be brought back to the original scaling. For instance, fixing $Q_0 = 341.3$, $\sigma = 5.67 \cdot 10^{-8}$, $K_\alpha = 0.1$, $K_\varepsilon = 0.5$, $T_\varepsilon = 288$, $\alpha_1 = \alpha_2 = 0.7$, $\varepsilon_1 = 0.7$, it can be shown that a cusp bifurcation occurs when $\varepsilon_2 \approx 0.6619$ for $\mu \approx 154.8768$ at $T \approx 288.035$. For lower values of $\varepsilon_2$, the bifurcation diagram $(\mu,T)$ has two saddle-node bifurcations; for higher values it has no saddle-node bifurcations. See Figure~\ref{fig:BD_emmisivity}.

\subsection{Butterfly catastrophe in the full system}

When $a>0$ and $c>0$, both albedo and emissivity change with temperature. In this case, the cusp bifurcations from both degenerate settings are present. In these bifurcations, there is a transition from two stable and one unstable equilibria to one stable equilibrium. Next to these, there is also another cusp bifurcation in the full system. In this additional cusp bifurcation, two unstable and one stable equilibria meet and become one unstable equilibrium. It is possible that these cusp bifurcations meet, and hence the five potential equilibria of the full system meet. This happens for parameter values for which $f_0 = f_1 = f_2 = f_3 = f_4 = 0$. Here, a bifurcation of codimension $4$ occurs, which is sometimes called a butterfly catastrophe.

Using the expressions found before, it is possible to find locations of butterfly catastrophes in the full system. For instance, fixing $c = 0.003$ and $d = 5$, using a numerical root finding algorithm we obtained the solution $\nu = 717271$, $a = 96965.4$, $y_r = 28.9109$, $z_\alpha = 0.212802$, $z_\varepsilon = -0.2214$.

Again, these results can be brought back to the original scaling. For this, we fix $Q_0 = 341.3$, $\sigma = 5.67 \cdot 10^8$, $K_\alpha = 0.1$, $K_\varepsilon = 0.5$, $\alpha_1 = 0.7$, $\varepsilon_1 = 0.7$. Then, by varying the remaining parameters it can be shown that a butterfly catastrophe occurs for $\varepsilon_2 \approx 0.6958$, $\alpha_2 \approx 0.4752$, $T_\alpha \approx 291.2370$, $T_\varepsilon \approx 286.8953$, $\mu \approx 143.0741$ and occurs at $T \approx 289.109$.

\begin{figure}

\begin{subfigure}[t]{0.45\textwidth}
	\centering
	\includegraphics[width=\textwidth]{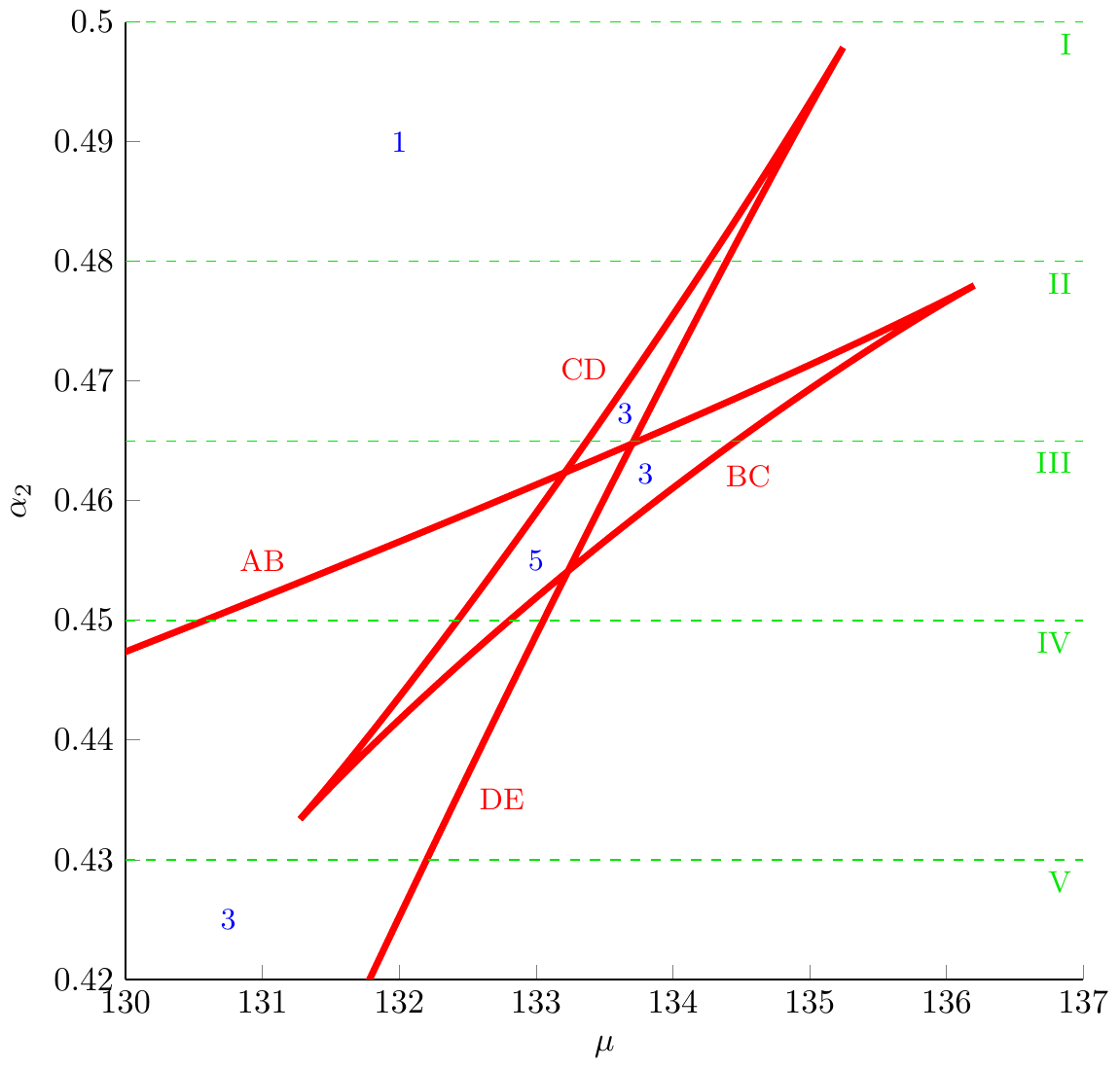}
	\caption{Lines of saddle-node bifurcations}
\end{subfigure}
\begin{subfigure}[t]{0.45\textwidth}
	\centering
	\includegraphics[width=\textwidth]{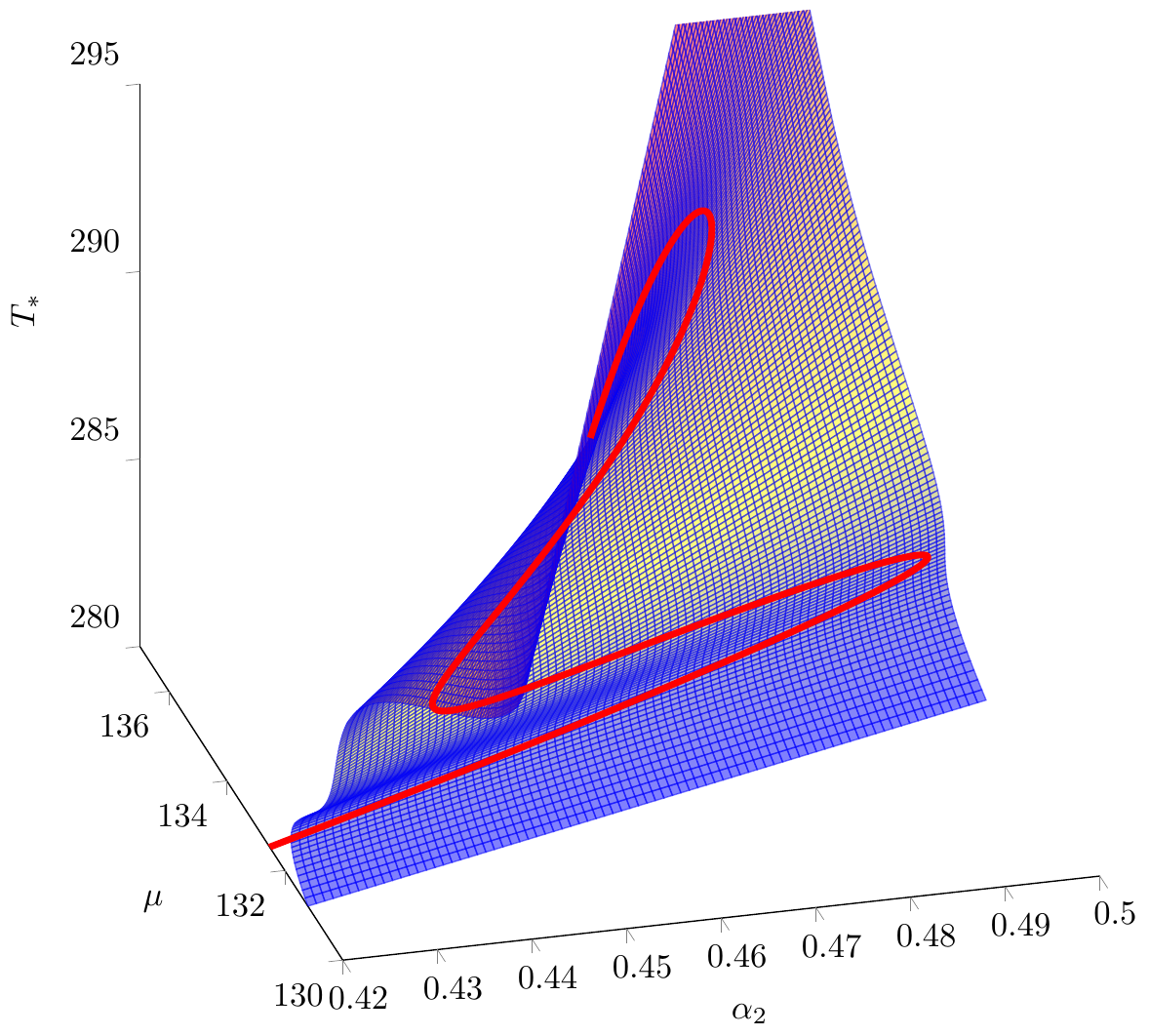}
	\caption{3D}
\end{subfigure}

\begin{subfigure}[t]{0.19\textwidth}
	\centering
	\includegraphics[width=\textwidth]{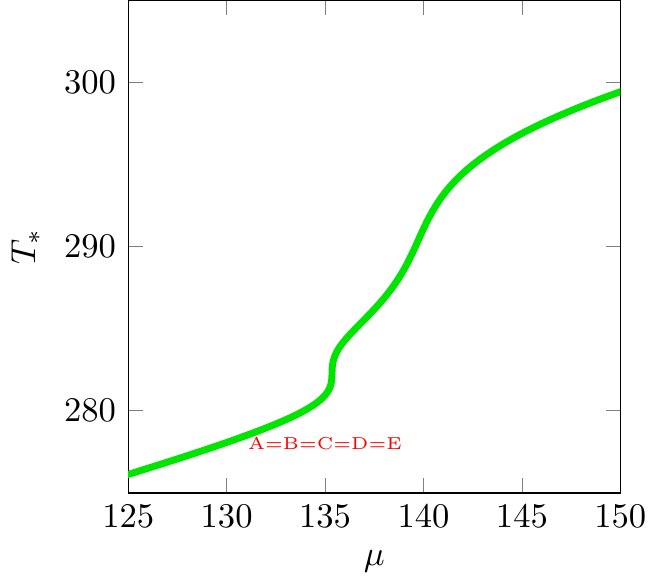}
	\caption{$\alpha_2 = 0.50$}
\end{subfigure}
\begin{subfigure}[t]{0.19\textwidth}
	\centering
	\includegraphics[width=\textwidth]{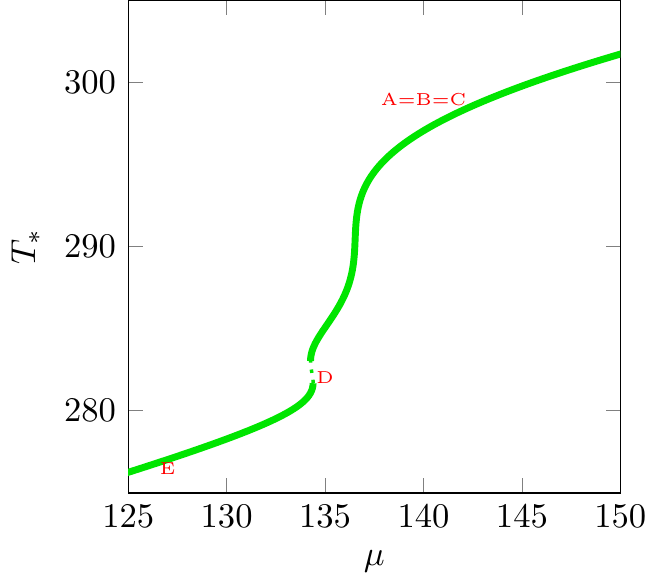}
	\caption{$\alpha_2 = 0.48$}
\end{subfigure}
\begin{subfigure}[t]{0.19\textwidth}
	\centering
	\includegraphics[width=\textwidth]{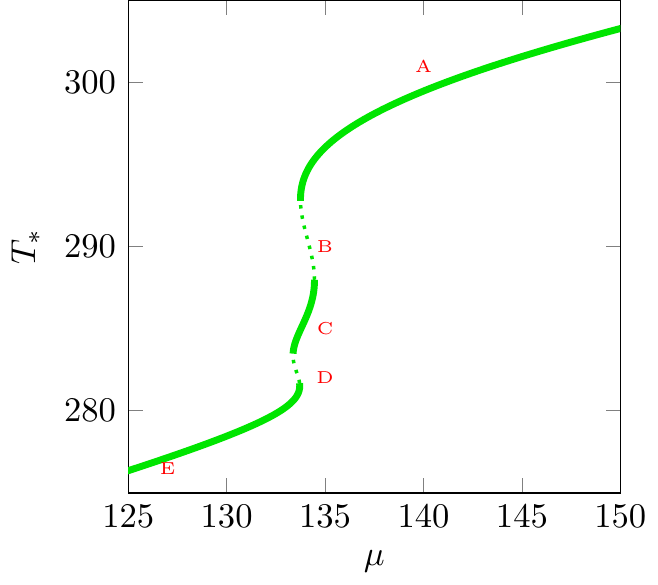}
	\caption{$\alpha_2 = 0.465$}
\end{subfigure}
\begin{subfigure}[t]{0.19\textwidth}
	\centering
	\includegraphics[width=\textwidth]{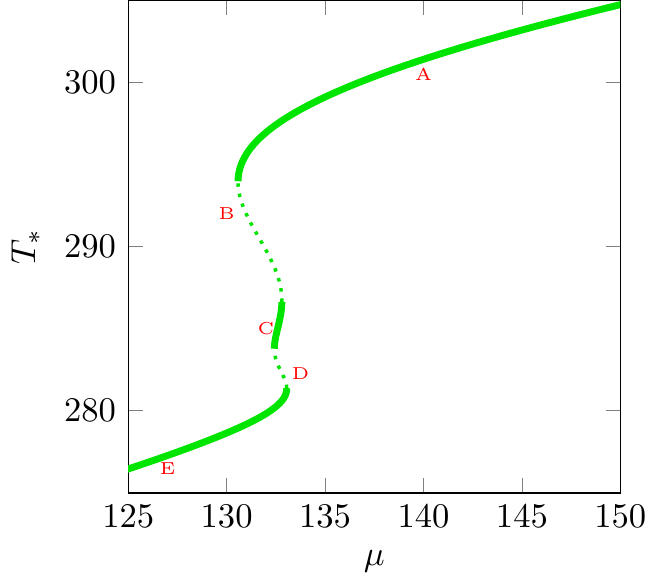}
	\caption{$\alpha_2 = 0.45$}
\end{subfigure}
\begin{subfigure}[t]{0.19\textwidth}
	\centering
	\includegraphics[width=\textwidth]{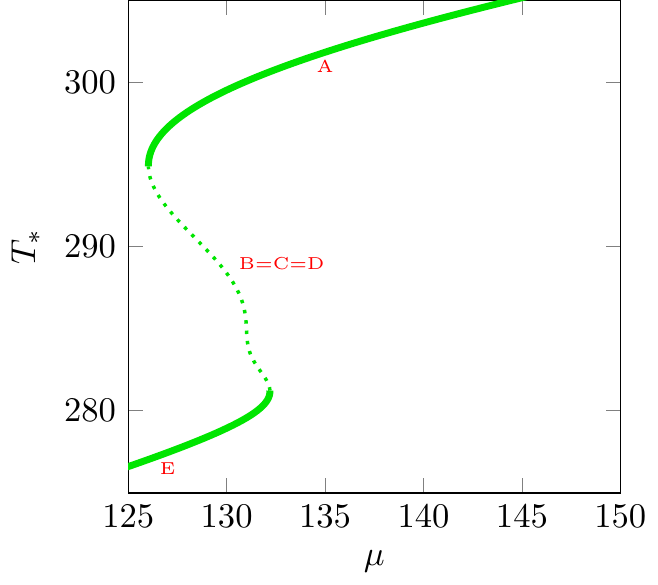}
	\caption{$\alpha_2 = 0.43$}
\end{subfigure}

\caption{Unfolding of the butterfly singularity. Here, parameters are taken close to the butterfly singularity and only parameters $\alpha_2$ and $\mu$ are varied in a numerical continuation made with Matcont~\cite{dhooge2008new}. (a) Loci of different fold points in the $(\mu,\alpha_2)$-parameter space. The blue numbers indicate the amount of fixed point in the different regions. The red letters indicate which equilibria meet up at the different fold curves, which correspond to the labels in figures (c-g). The label is located at the side of the fold branch where the equilibria have merged. (b) Surface of equilibria points for different values of $\alpha_2$ and $\mu$. The red curve denotes the loci of fold points. (c-g) Bifurcation diagrams $(\mu,T_*)$ for different fixed values of $\alpha_2$, corresponding to the different green dashed lines in (a). In these diagrams dotted lines indicate unstable equilibria, and solid lines stable equilibria. The red letters are labels of the different equilibria, which correspond to the labels of the fold curves in (a). In these computations, parameter values are $Q_0 = 341.3$, $\sigma = 5.67 \cdot 10^{-8}$, $K_\alpha = 0.1$, $K_\varepsilon = 0.5$, $\alpha_1 = 0.7$, $\varepsilon_1 = 0.7$, $T_\alpha = 291$, $T_\varepsilon = 282$ and $\varepsilon_2 = 0.68$.}
\label{fig:BD_butterfly}
\end{figure}

For parameters close to this point, the degeneracy unfolds into four saddle-node branches, with three cusp points. In Figure~\ref{fig:BD_butterfly}, we show such a unfolding where we follow the fold loci as parameters $\alpha_2$ and $\mu$ vary, and all the other parameters are taken close to above found butterfly singularity. In the diagrams the three cusps are located where the fold curves meet. If parameters would be taken closer to the butterfly singularity, these cusps points move together and meet up precisely at the singularity.

\clearpage

\newpage

\section{Supplementary material}

\subsection{Region tipping in the CESM 1.0.4 abrupt8xCO2 run}
\label{app:CESM8xCO2}

\begin{figure}[hb!]
	\centering
	\begin{subfigure}[t]{0.45\textwidth}
		\includegraphics[width=\textwidth]{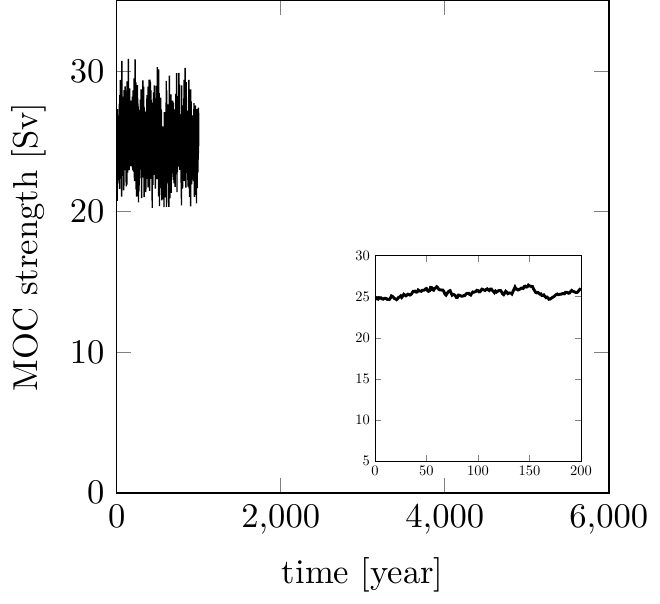}
		\caption{control experiment}
	\end{subfigure}
	\begin{subfigure}[t]{0.45\textwidth}
		\includegraphics[width=\textwidth]{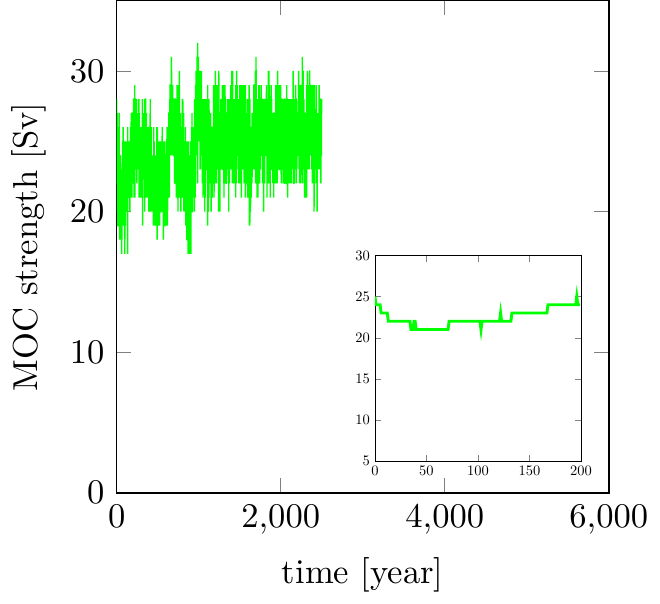}
		\caption{abrupt2xCO2 experiment}
	\end{subfigure}
	\begin{subfigure}[t]{0.45\textwidth}
		\includegraphics[width=\textwidth]{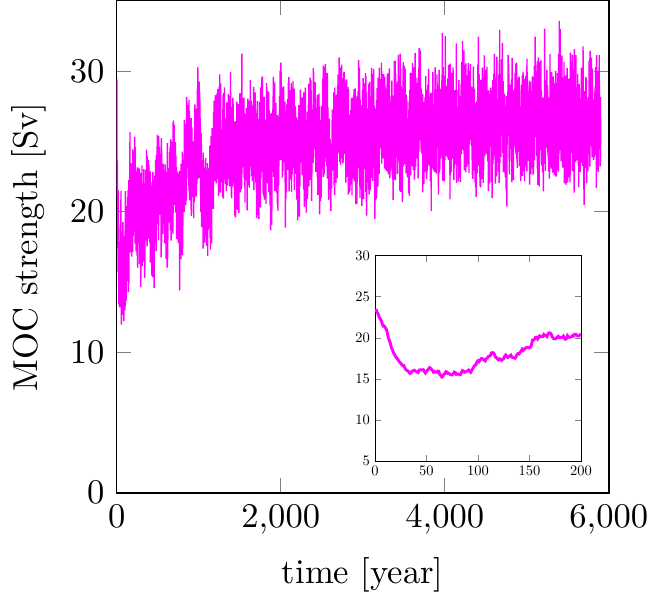}
		\caption{abrupt4xCO2 experiment}
	\end{subfigure}
	\begin{subfigure}[t]{0.45\textwidth}
		\includegraphics[width=\textwidth]{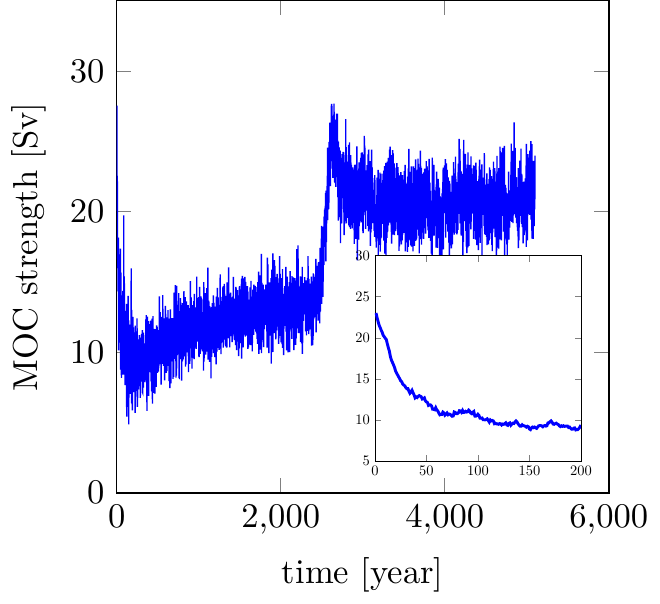}
		\caption{abrupt8xCO2 experiment}
	\end{subfigure}

	\caption{Strength of the (global) meridional overturning circulation in control (a), abrupt2xCO2 (b), abrupt4xCO2 (c) and abrupt8xCO2 (d) experiments in the CESM 1.0.4. contribution to longrunMIP~\cite{rugenstein2019longrunmip}. This strength is measured as the (yearly averaged) maximum of the global stream function of depths below $500m$ in the Northern Hemisphere. The insets show a $50$ year rolling average for the first $200$ years of each experiment. In all abrupt \COO\ forcing experiments, a weakening can be seen at the start of the runs. In the 2xCO2 and 4xCO2 experiments, this is restored gradually; in the 8xCO2 experiment, the AMOC lingers in a weakened state, and suddenly restores around $t = 2400$ years.
	}
	\label{fig:longrunmip-AMOC}
\end{figure}

Figure~\ref{fig:longrunmip-AMOC} shows that the abrupt addition of atmospheric \COO\ causes a rapid weakening of the AMOC in the model CESM 1.0.4. In the abrupt2xCO2 and abrupt4xCO2 experiments this weakening gets restored gradually over time, but in the abrupt8xCO2 the system lingers around in a weakened state for long and then suddenly restores rapidly around $t = 2400$ years, which seem to be the cause of the rapid increase in global warming this late in the run.

\subsection{Expressions for $f_i$.}
\label{sec:fi}

The expressions for $f_i$ in (\ref{eq:f}) can be computed as:
$$
\begin{aligned}
f_0 := &\ \nu - a \tanh(z_\alpha) - y_r^4 - c y_r^4 \tanh(d z_\varepsilon)\\
f_1 := &\ a \sech^2(z_\alpha) - 4 y_r^3 + c d y_r^4 \sech^2(d z_\varepsilon) - 4 c y_r^3 \tanh(d z_\varepsilon)\\
f_2 := &\ a \tanh(z_\alpha)\sech^2(z_\alpha) - 6 y_r^2 + cd^2 y_r^4 \tanh(d z_\varepsilon)\sech^2(d z_\varepsilon) + 4 cd y_r^3 \sech^2(d z_\varepsilon) - 6 c y_r^2 \tanh(d z_\varepsilon)\\
f_3 := &\ \frac{a}{3} \sech^2(z_\alpha)\left[2 \tanh^2(z_\alpha) - \sech^2(z_\alpha)\right] - 4 y_r + \frac{cd^3}{3} y_r^4 \sech^2(d z_\varepsilon) \left[2 \tanh^2(d z_\varepsilon) - \sech^2(d z_\varepsilon)\right]\\
&\ + 4 c d^2 y_r^3 \tanh(d z_\varepsilon)\sech^2(d z_\varepsilon) + 6 c d y_r^2 \sech^2(d z_\varepsilon) - 4 c y_r \tanh(d z_\varepsilon) \\
f_4 := &\ - \frac{a}{3} \tanh(z_\alpha)\sech^2(z_\alpha) \left[ 2\sech^2(z_\alpha) - \tanh^2(z_\alpha) \right] - 1\\
&\ - \frac{cd^4}{3} y_r^4 \tanh(d z_\varepsilon)\sech^2(d z_\varepsilon)\left[2 \sech^2(d z_\varepsilon) - \tanh^2(d z_\varepsilon)\right] \\
&\ + \frac{4cd^3}{3} y_r^3 \sech^2(d z_\varepsilon)\left[2\tanh^2(d z_\varepsilon)-\sech^2(d z_\varepsilon)\right] + 6 cd^2 y_r^2 \tanh(d z_\varepsilon)\sech^2(d z_\varepsilon) \\
&\ + 4 cd y_r \sech^2(d z_\varepsilon) - c \tanh(d z_\varepsilon) \\
f_5 := &\ \frac{a}{15} \sech^2(z_\alpha) \left[2 \sech^4(z_\alpha) + 2 \tanh^4(z_\alpha) - 11\sech^2(z_\alpha)\tanh^2(z_\alpha)\right] \\
&\ + \frac{cd^5}{15} y_r^4 \sech^2(d z_\varepsilon)\left[2 \sech^4(d z_\varepsilon) + 2 \tanh^4(d z_\varepsilon) - 11 \sech^2(d z_\varepsilon)\tanh^2(d z_\varepsilon)\right] \\
&\ - \frac{4 c d^4}{3}y_r^3 \tanh(d z_\varepsilon)\sech^2(d z_\varepsilon) \left[2  \sech^2(d z_\varepsilon) - \tanh^2(d z_\varepsilon) \right] \\
&\ + \frac{6 c d^3}{3} y_r^2 \sech^2(d z_\varepsilon)\left[2\tanh^2(d z_\varepsilon)-\sech^2(d z_\varepsilon)\right] + 4 c d^2 y_r \tanh(d z_\varepsilon)\sech^2(d z_\varepsilon) \\
&\ + cd \sech^2(d z_\varepsilon)
\end{aligned}
$$

\end{document}